\documentclass{article}

\usepackage[english]{babel}
\usepackage{setspace}
\usepackage[capposition=top]{floatrow}
\usepackage[bottom]{footmisc}
\usepackage[letterpaper,top=2cm,bottom=2cm,left=3cm,right=3cm,marginparwidth=1.75cm]{geometry}

\usepackage{booktabs}
\usepackage{graphicx} 
\usepackage[colorlinks=true, allcolors=blue]{hyperref}
\usepackage[authoryear]{natbib}
\usepackage{amsmath}
\usepackage{threeparttable}
\usepackage{caption}
\usepackage{subcaption}
\usepackage{longtable}
\usepackage{array}
\usepackage{caption}
\usepackage{colortbl}
\usepackage{color,soul}

\usepackage{siunitx}
\DeclareMathOperator*{\argmin}{arg\,min}

\usepackage{float}
\floatstyle{plaintop}
\restylefloat{table}

\begin{document}

\title{Fiscal Spillovers through Informal Financial Channels\thanks{I thank my advisor, Francisca Antman, for her enduring support and direction throughout this project. I also thank Murat Iyigun, Tania Barham, Brian Cadena, Steven Billings, Clemens Graf von Luckner, Rimjhim Saxena, Vincent Mastantuno, Mark Spiegel and two anonymous referees, as well as the participants of the applied microeconomics and graduate student brownbags at the University of Colorado, Boulder for incredibly helpful comments. All errors are my own.}}

\author{Austin Kennedy\footnote{University of Colorado, Boulder; \href{mailto:austin.kennedy@colorado.edu}{austin.kennedy@colorado.edu}}}
\date{June 8, 2025\\
\vspace{1em}
Accepted Manuscript\footnote{\textcopyright\ Austin Kennedy. This accepted manuscript is made available under the CC BY-NC-ND 4.0 license. To view a copy of this license, visit \href{https://creativecommons.org/licenses/by-nc-nd/4.0/}{https://creativecommons.org/licenses/by-nc-nd/4.0/}}\\
\vspace{1em}
First Draft: April 2024\\
\vspace{1em}
Revised: March 2025\\
\vspace{1em}
Accepted: June 2025\footnote{This is the accepted manuscript of the article: ``Fiscal Spillovers through Informal Financial Flows." \textit{Journal of International Money and Finance,} 157: 1033-78, August 2025. The final published verison is available at  \href{https://doi.org/10.1016/j.jimonfin.2025.103378}{https://doi.org/10.1016/j.jimonfin.2025.103378} \textcopyright\ Elsevier 2025}\\
}

\maketitle

\begin{abstract}
\singlespacing\noindent
    This paper examines fiscal policy spillovers through informal international financial channels, using the US stimulus checks as a positive, sudden, and direct fiscal shock. I utilize granular, transaction-level cryptocurrency data combined with an algorithm to probabilistically identify cross-border ``crypto vehicle" transactions to construct bilateral cryptocurrency flows between countries. Using a difference-in-differences strategy, I compare cryptocurrency outflows between the US and other high-income countries and find a sharp but temporary increase in cryptocurrency outflows as a result of the direct stimulus. I quantify the fiscal spillover relative to expenditure and place an upper bound of 2.52\% through this channel. This implies that fiscal spillovers through remittance channels are likely modest in size.

\thispagestyle{empty}

\bigskip \singlespace \noindent


\bigskip

\pagebreak
\end{abstract}

\renewcommand{\thefootnote}{\arabic{footnote}}
\setcounter{page}{1}

\section{Introduction}

Times of crisis often call for swift, decisive, and effective action to be taken by policymakers. The COVID-19 pandemic is a poignant example of such a crisis and subsequent policy response, with most countries around the world taking drastic action to slow the virus's spread and to keep their economies afloat. The United States (US) in particular enacted the single largest fiscal stimulus in its history in response to the pandemic--the CARES Act--totaling \$2.3 trillion, or $\sim$9\% of its GDP, including \$271 billion in direct, lump-sum payments to individuals.

In this paper, I ask the question: what was the effect of the US direct fiscal stimulus on outward remittances? This financial channel has been largely unexplored by the existing literature on fiscal spillovers, yet has a large potential to dampen the ability of policymakers to respond to a crisis with fiscal stimulus, if remittances allow that stimulus to exit the domestic economy and avoid taking part in the virtuous positive feedback loop associated with a fiscal expansion.

Investigating this question empirically has historically proven difficult due to the available data--most macroeconomic data on remittances is highly aggregated and only available at relatively infrequent intervals, typically measured annually. I overcome this issue by exploiting data from a relatively new and under-explored remittance channel--cryptocurrency. With transaction-level data from the peer-to-peer (P2P) cryptocurrency exchange Paxful, I use an algorithm proposed by \cite{decrypting2023} to probabilistically identify cross-border cryptocurrency transactions, which I then use to construct a dataset of bilateral cryptocurrency flows. As transactions are measured down to the second, this granularity allows for much tighter identification of fiscal spillovers around fiscal policy shocks, in this case the Economic Impact Payments (EIPs), or stimulus checks, that were disbursed by the US government as part of the CARES Act.

{Using a difference-in-differences strategy which exploits the timing of the disbursement of stimulus checks to compare the trajectory of cryptocurrency outflows between the US and cryptocurrency outflows of other countries, I find evidence of a sharp but temporary increase in cryptocurrency outflows due to the disbursement of EIPs, with an overall increase of around 11\%, and higher increases of  17\% and 16\% found for cryptocurrency destined to middle- and high-income countries, respectively. No effect is found for cryptocurrency destined to low-income countries. Using the estimates from my model, I quantify the overall fiscal spillover through this channel to be a roughly 1.27\% increase in total cryptocurrency outflows from the US for the year 2020 due to the disbursement of EIPs. This quantification allows me to extrapolate the fiscal spillover to other financial channels that are measured annually, assuming a similar response to the disbursement of EIPs. I perform a back-of-the-envelope calculation to put a bound on the size of fiscal spillovers relative to total expenditure, finding that fiscal spillovers are unlikely to exceed 2.5\% of total spending on EIPs, and more likely fall in the range of 0.2-0.7\% of expenditure.

The modest size of this estimate suggests that fiscal spillovers through this channel are unlikely to dampen the effectiveness of direct fiscal stimulus. Other concerns, such as speed of disbursement, identity fraud, or the targeting of fiscal stimulus towards those with a higher marginal propensity to consume (MPC) should be prioritized by policymakers. That being said, a small financial outflow from the perspective of the US may still represent a large inflow for a small, remittance-receiving country. Policymakers in these countries may wish to mitigate or amplify their exposure to the fiscal actions of others depending on their policy goals. For example, these fiscal spillovers offer a chance to ``free ride" off of the fiscal policy of another country, which may be desirable for a country facing a negative economic shock along with constraints to their ability to respond. On the other hand, for a country experiencing inflation, exposure to the fiscal stimulus of other countries may exacerbate the problem, thus policy to block channels for fiscal spillovers may be warranted.

This paper adds to the evidence on the effectiveness of direct fiscal policy in responding to economic crises. This literature has largely focused on the MPC of direct vs. indirect fiscal actions, and the state-dependence of the MPC depending on the type of negative shock an economy is experiencing. Studies such as \cite{sahm2012}, \cite{parker2013}, and \cite{broda2014} find evidence that MPCs are higher for large, lump-sum stimulus payments as opposed to other forms of fiscal policy, and that the consumption response to direct fiscal stimulus is rapid, often occurring within a few months of its receipt.\footnote{See \cite{sahm2019direct} for a larger discussion about direct vs. indirect stimulus and the existing evidence.} More recently, \cite{cox2020}, \cite{baker2023}, and \cite{chetty2023} all find a strong consumption response to the disbursement of EIPs during the COVID-19 pandemic, especially for lower-income households. Despite this, there is also evidence that fiscal stimulus may be less effective at stimulating demand in the presence of supply-side shocks, such as those present during the pandemic. \cite{ramey2018} and \cite{brunet2024} suggest that fiscal multipliers were lower during WWII as a result of restrictions on household spending. \cite{auerbach2022} study this in the context of COVID-19, finding that supply-side shocks such as pandemic lockdowns may dampen the effectiveness of fiscal policy in stimulating demand, by decreasing the ability of the economy to absorb government spending into new production and employment. \cite{ghassibe2022} additionally find that fiscal stimulus during a supply-side shock may crowd out private consumption and even potentially generate a negative multiplier. The present study adds to this debate by examining a new channel that could dampen the effectiveness of fiscal policy, although my results suggest that any negative impacts from this channel are likely modest.

This study also contributes to the literature on fiscal spillovers. The direction of fiscal spillovers and the channel through which they operate is still debated. Most studies on the subject focus on the effect that a fiscal expansion from a large economy has on real exchange rates and real interest rates, and the subsequent impact on output for foreign economies. The theoretical results suggested by standard real business cycle models or old and new-Keynesian theories imply that expansionary US policy should lead to an increase in the real exchange rate, thus decreasing the trade balance and increasing the output of its trade partners. However, most of the empirical time-series evidence suggests precisely the opposite, that real exchange rates decrease as a result of a fiscal expansion, thus increasing net exports for the US and decreasing output for US trade partners.\footnote{see \cite{ferrara_questioning_2021} for an excellent overview of the existing literature.} More recently, \cite{auerbach2013} and \cite{ferrara_questioning_2021} find an increase in real exchange rates as a result of fiscal expansion, using unexpected changes in military expenditure combined with SVAR models to identify the relationship. On the other hand, \cite{faccini_international_2016} suggests that fiscal spillovers operate through the effect of fiscal expansions on real interest rates instead. They argue that a fiscal expansion in the US decreases real interest rates, ultimately increasing output in its trade partners as a result. Finally, \cite{KUMAR2024103168} argue that, for emerging markets, the impact of fiscal policy in the US on emerging market economies is negative, as the impact of a depreciation in the exchange rate and subsequent decrease in the demand for exports from emerging economies dominates the decrease in real interest rates caused by expansionary fiscal policy in the US. I contribute to this literature by exploring remittances as a potential channel for fiscal spillovers, which has been largely unexplored by the existing literature. Additionally, most of the literature on fiscal spillovers studies the effect that a domestic fiscal expansion has on foreign economies, rather than studying the implications of fiscal spillovers for the country performing the stimulus. I provide evidence that, while fiscal spillovers through remittances are likely present, they are relatively small compared to overall expenditure on fiscal stimulus, and are thus unlikely to significantly hamper policymakers' ability to respond to an economic crisis with fiscal policy.

This paper additionally adds to the literature on remittances in the context of the COVID-19 pandemic. Several studies, such as \cite{cardozo_silva_impact_2022}, \cite{carare_evolution_2022}, \cite{kpodar_defying_2023}, and \cite{bansak2025} document the initial dip in remittances due to the onset of COVID-19 and the rapid subsequent recovery in the second half of 2020, using officially recorded remittance data. \cite{kpodar_defying_2023} and \cite{bansak2025} both document a positive relationship between remittances and the enactment of fiscal stimulus in migrant host countries, using monthly data collected from central banks on formal remittances. My paper adds to this evidence, as I am able to construct bilateral cryptocurrency flows at an even higher frequency and for all corridors, allowing me to causally identify the impact of EIPs on cryptocurrency flows.  My findings also suggest that, in this context at least, informal remittances show similar behavior to those flowing through traditional channels such as money operators or banks. This is a signficant finding given that informal remittances are estimated to be anywhere between 50 to 250 percent of officially recorded remittances.\footnote{Estimates of the size of informal remittances come from \cite{freund2008}, \cite{ratha2011}, and \cite{amjad2013}.} Finally, my results are consistent with the finding of \cite{kpodar_defying_2023} of a shift away from remittances delivered via non-digital means (e.g. cash carried via travel) and towards digital payments instead.

Finally, this paper contributes to a nascent and growing literature on cryptocurrency. Much of the existing cryptocurrency literature examines whether or not cryptocurrency satisfies the characteristics of money (e.g. \cite{baur2021volatility}, \cite{ardia2019regime}, \cite{bouri2017does}), with most coming to the conclusion that it does not.\footnote{One exception is \cite{ueda2021}, which finds evidence of price stability in cryptocurrency relative to other major financial assets.} Some studies have documented cryptocurrency usage, mostly in regard to usage for illegal activity (\cite{foley2019sex}, \cite{cong2022crypto}). This paper studies a widely discussed but ill-studied use-case for cryptocurrency as a cross-border financial vehicle. This paper closely relates to \cite{decrypting2023}, who develop the algorithm I use to identify cross-border cryptocurrency transactions, and describe some basic facts about cryptocurrency's usage as a financial vehicle, but stop short of any causal identification. I go beyond their analysis by examining the response of cross-border cryptocurrency flows to a wealth shock. Finally, my paper is also closely related to \cite{divakaruni_uncovering_2021}, who study the impact of the US COVID stimulus checks on retail trading in Bitcoin and find an increase in retail trading as a result of the stimulus. This paper extends their analysis by focusing on cross-border crypto-vehicle transactions rather than the overall volume of cryptocurrency trading. Altogether, I contribute to this literature by documenting a new and potentially important use-case for cryptocurrency, along with potential implications policymakers both in countries performing fiscal stimulus and for countries potentially exposed to the fiscal stimulus of others through migrant networks.

The paper proceeds as follows. Section \ref{sect:Data} details the data used in my analysis as well as the transaction matching algorithm used to construct cross-border cryptocurrency flows, Section \ref{sect:empirics} lays out my empirical strategy as well as some background on the US stimulus checks, Section \ref{sect:results} presents the results of my main analysis, Section \ref{sect:discussion} provides an estimation of the overall spillover effect relative to expenditure, and discusses the policy implications of my findings, Section \ref{sect:robustness} shows a couple of robustness checks, and Section \ref{sect:conclusion} concludes.

\section{Data}\label{sect:Data}

\subsection{Paxful}\label{sect:data_paxful}

I use data from the centralized ``off-chain" cryptocurrency exchange Paxful.\footnote{``On-chain" cryptocurrency refers to cryptocurrency bought and sold on the blockchain, which is a decentralized public ledger of transactions. This is as opposed to ``off-chain" cryptocurrency, such as the data that I use, that occurs through centralized cryptocurrency exchanges.} I scraped data through their Application Programming Interface (API), covering the universe of transactions that occurred on Paxful from March 2017 to September 2022. Each observation represents a transaction, and contains a unique transaction ID, a timestamp measured to the second, data on the size of the transaction in Bitcoin measured down to the santosh (eight decimal places), the fiat currency used to buy Bitcoin, the price paid (fiat currency per one unit of Bitcoin), as well as the geolocation (at the country level) of the user and advertiser involved in the transaction and the payment method (e.g. bank transfer, wire transfer, gift card, etc.).

These data are ideal for my research question due to Paxful operating in every country around the world and being lightly regulated, thus lending itself to informal finance.\footnote{This is due to the fact that Paxful only holds user deposits in Bitcoin, not fiat currency. Exchanges of fiat are made in a peer-to-peer fashion, hence individual buyers/sellers are the holders of fiat, while Paxful only holds Bitcoin, thus avoiding regulation that some other cryptocurrency exchanges are subject to.} Additionally, cryptocurrency exchanges like Paxful are ideal for international financial transactions, especially in countries with lower levels of financial infrastructure, as one can avoid the high fees and delays associated with traditional bank transfers or remittances sent via money transfer operators such as Western Union and MoneyGram, as well as avoiding capital controls. Paxful and similar exchanges also make it relatively easy to convert fiat currency into cryptocurrency while avoiding the formal financial system. For example, instead of linking one's bank account to Paxful in order to purchase or sell cryptocurrency, one could instead exchange their Bitcoin for a gift card to various retail chains or to an electronic platform such as iTunes, or a prepaid debit card.

\begin{figure}[!h]
    \centering
    \includegraphics[scale=0.6]{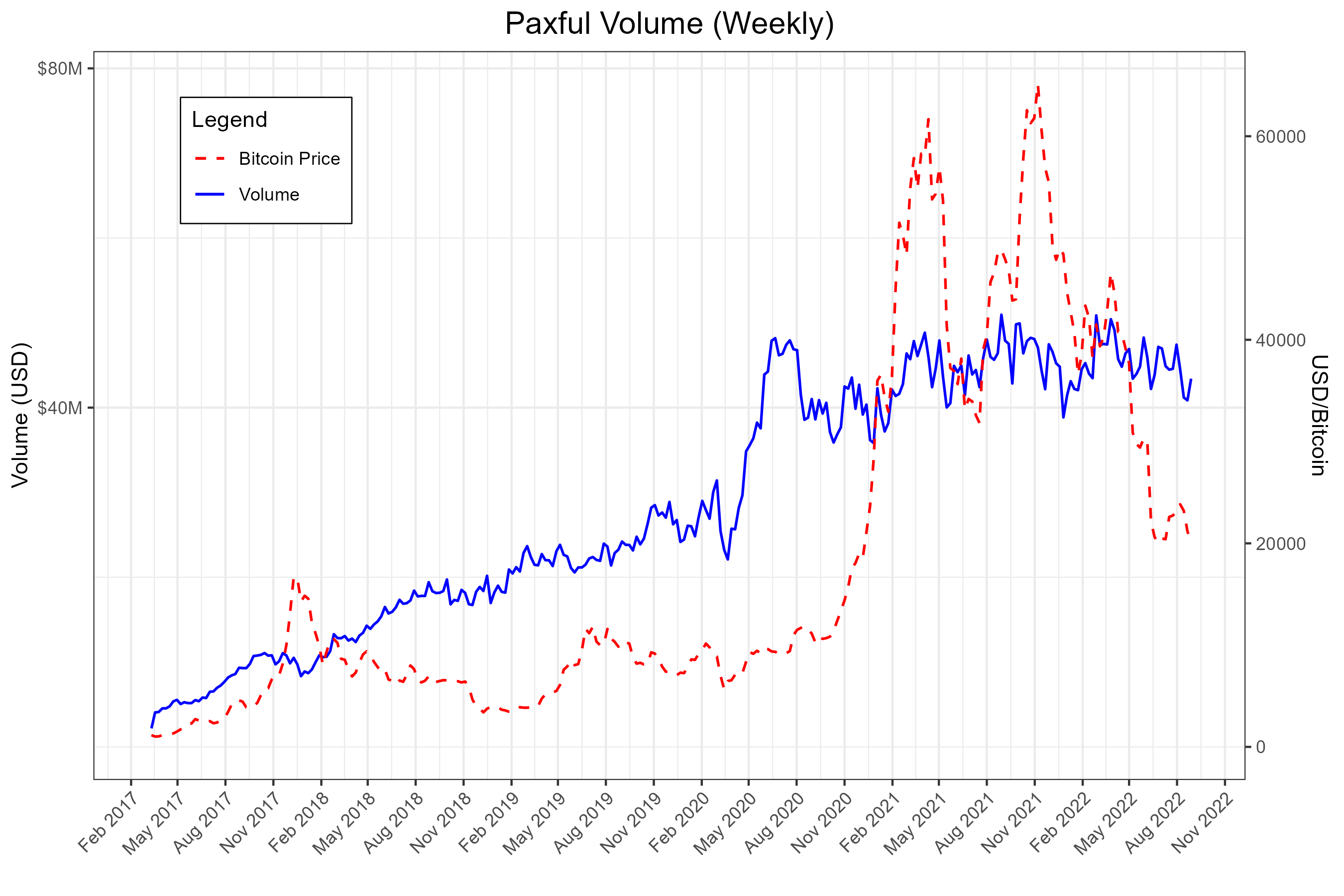}    \floatfoot{Source: Paxful, Author's calculations}
    \caption{Paxful Volume and USD/Bitcoin Exchange Rate}
    \label{fig:weekly_volume}
\end{figure}

I plot the total weekly volume, measured in USD, of all cryptocurrency transactions on Paxful as well as the USD/Bitcoin exchange rate in Figure \ref{fig:weekly_volume}. This figure shows that the total volume of transactions occurring on Paxful steadily increases throughout the period for which I have data, in line with broader cryptocurrency usage throughout this timeframe. Unsurprisingly, the total volume of transactions on Paxful appears to be impacted by macroeconomic shocks. Notably, once COVID lockdowns begin in early 2020 there is a dip in transaction volume, followed by a sharp, temporary spike once governments around the world intervened and fiscal stimulus led to an increase in disposable income for consumers. Importantly, for my later analysis studying the response of international cryptocurrency flows to the US COVID stimulus checks, the relative exchange rate of USD for Bitcoin is relatively stable throughout 2020, with a speculative bubble occurring later in 2021. This makes it more likely that what I am identifying truly is an increase in international transfers, rather than speculative behavior.

While I do not have any data on user characteristics in this data, to shed light on the potential for selection bias it is useful to try and characterize the type of individual who may tend to use a platform such as Paxful. Generally the literature on the demographics of cryptocurrency usage is focused on retail traders in developed countries, finding that they tend to be young, male, single, technologically skilled, and earning higher income.\footnote{See \cite{AUER2022101066} and \cite{english2020}, for example.} To my knowledge, there is no evidence regarding the demographics of cryptocurrency users in less developed countries or of those using cryptocurrency as a financial vehicle rather than as a speculative asset.

In my setting, where users intend to use cryptocurrency as a financial vehicle, users would need to be relatively competent with technology in order to set up an account and use Paxful to make transactions, thus making the average user likely younger rather than older. The underlying technological infrastructure required to use a platform such as Paxful is not very limiting, however, as one really only needs an internet connection and a device on which to use Paxful such as a cellphone or a computer. As mentioned previously, one need not be connected to the formal financial system to make transactions on Paxful. This indicates that, even in lower income countries with less financial infrastructure, at least some portion of the population has the potential to be users. When I limit my data to exclusively international transactions in my main analysis, this will require that a user has a desire to send money internationally and has reasons to avoid the formal international financial system, be it avoiding high fees, delays, capital controls, or a desire to remain anonymous. All in all, these data are likely skewed towards younger users residing in countries in the ``sweet spot" in terms of the potential benefit of using cryptocurrency, where the technological infrastructure is developed enough to support cryptocurrency usage, but financial institutions are underdeveloped enough to incentivize cryptocurrency usage for international transactions.

Relating this discussion back to the internal validity of my empirical strategy, I implicitly need to assume that the demographics of users on Paxful stayed relatively constant around the disbursement of stimulus checks, insofar as changes in the cryptocurrency outflows from the US do not simply reflect new users who tend to send more money internationally joining the platform around the time of stimulus check disbursement. I cannot explicitly test this assumption given the anonymous nature of transacting on Paxful and lack of user characteristics or even unique user IDs in the data. I argue that this assumption is plausible, however, due to the fact that broad public interest in cryptocurrency did not begin to gain momentum until the end of 2020 and beginning of 2021, as shown by the large increase in Bitcoin price in Figure \ref{fig:weekly_volume}. Figure \ref{fig:weekly_volume} additionally shows that the overall volume of cryptocurrency transactions rises at a relatively constant rate throughout the time for which I have data. If the volume of cryptocurrency transactions on Paxful suddenly and permanently spiked around the disbursement of stimulus checks, I would be more concerned that a changing user base could be driving my results. Instead, what I observe is a short-lived spike in cryptocurrency volume around the disbursement of stimulus checks, but no permanent increase in the trend in cryptocurrency volume following this period.\footnote{A Zivot-Andrews test indicates a structural break around the beginning of August 2020, however the trend in cryptocurrency volume after this point slowly decreases. As this is outside of the range of my main analysis and a permanent \textit{increase} in trend is what I would be concerned about, I do not consider this to be a threat to internal validity.}

A limitation of this paper is that data from Paxful may not be representative of cryptocurrency as a whole, or of informal international finance more broadly. While I acknowledge this limitation, I argue that these data are the best look into these opaque and difficult to measure markets that is currently publicly available. I also argue that, while the opaqueness of cryptocurrency as a market as well as the platforms that support its usage make Paxful's representativeness to cryptocurrency as a whole unknowable, the characteristics that make Paxful favorable for international financial transactions are true of the broader cryptocurrency market as well. These characteristics include the anonymous nature of transacting on Paxful, that transactions are instantaneous and cheap relative to formal international financial channels, and available at all times of day and days of the week. I show a comparison of Paxful's size to the top five cryptocurrency platforms, the cryptocurrency market as a whole, a few major remittance platforms, and officially recorded remittance flows in Table \ref{tab:volume_comparison}. Given the decentralized nature of the cryptocurrency market, it is unsurprising that Paxful comprises a negligible share of the market. Even Binance, the largest cryptocurrency platform in 2020, captured less than 3\% of cryptocurrency's total volume on its platform. Given the level of fragmentation in the cryptocurrency market, representative data likely does not exist, and I again argue that this is the best publicly available data on cryptocurrency usage, especially as it relates to international transfers.

Even if Paxful is representative of cryptocurrency's usage for international finance, there is still the question of the external validity of this study with regards to international finance as a whole. For the effect of the disbursement of stimulus checks on cryptocurrency outflows from the US to capture the overall pattern of fiscal spillovers, I would need to assume that the global pattern of international financial flows are well represented by those identified on Paxful. Given the particular characteristics of cryptocurrency users, and especially those intending to use cryptocurrency as an international financial vehicle, I find Paxful less likely to be representative of formal international financial markets than of cross-border cryptocurrency transactions as a whole. International financial transactions through traditional channels are conducted by a wide array of individuals, firms, and institutions with a variety of reasons for participating in international finance, and many of these actors are unlikely to use or benefit from cryptocurrency. Still, as informal financial flows are difficult to measure and are excluded from most data on international finance, Paxful may offer a glimpse into this under-studied component of international finance than what has been done in previous work. In any case, in Section \ref{sect:discussion} I provide estimates of the total size of fiscal spillovers under various assumptions about the generalizability of my findings to the cryptocurrency market, the formal remittance market, and the informal remittance market.

\subsection{Matching Algorithm}

I follow the methodology of \cite{decrypting2023} to probabilistically discover ``crypto-vehicle" trades in the data. This methodology makes use of the granularity of the cryptocurrency data, whereby the amount of the trade in Bitcoin is measured down to the santosh (eight decimal places), and the time of each trade is measured down to the second. When two trades of exactly equal size are observed within a given time window, a p-value that this match is non-random can be assigned, using the past distribution of Bitcoin trade sizes. Specifically, for a given trade $i$ with trade size $x_{i}$ this p-value is given by the following:\footnote{See \cite{decrypting2023} for a derivation and full explanation.}

\begin{equation}
    1 - (1-p_{i})^{N_{i}}
\end{equation}

Where $p_{i}$ is the probability of observing a trade of size $x_{i}$ given the distribution of trade sizes up until trade $i$ occurs, and $N_{i}$ is the number of trades that occur within the time window of interest after the first trade occurs. For the purposes of this paper, I follow \cite{decrypting2023} in using a five hour window and a threshold of 0.05 as the maximum p-value for a trade to be classified as a crypto-vehicle trade.

This algorithm accounts for the possibility that two trades of equal size may randomly occur, hence leading to a false positive that two equal transactions are being used for as a financial vehicle. Using the prior distribution of trades to calculate $p_{i}$ accounts for the fact that some Bitcoin trade sizes are simply more popular than others, for example 0.10 Bitcoin. For the majority of matched trades in the data, the trade size is unique, i.e. there are only two trades that take on a certain trade size within the entire dataset, and they occur within a short time window.

A convenient feature of cryptocurrency that lends itself to the discovery of financial vehicle trades is the volatility of its price measured in fiat currency. Even if certain fiat trade sizes, for example \$10, are relatively common, these trade sizes will change from one second to the next when measured in Bitcoin. Hence, when two trades, measured in Bitcoin, of the exact same size occur within a short window, the probability that they represent a financial vehicle transaction is high.

\begin{figure}[!h]
    \centering
    \includegraphics[scale = 0.5]{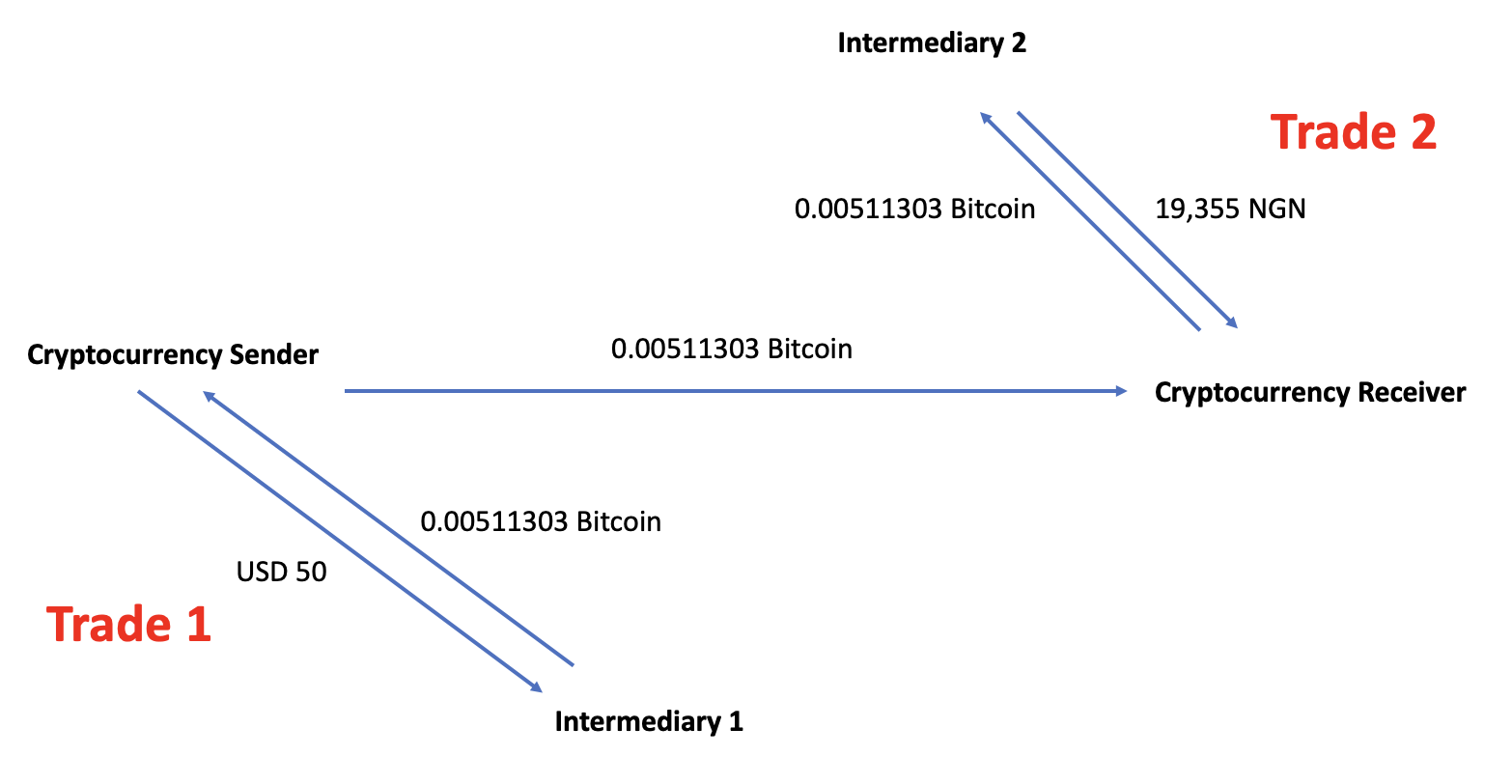}
    \caption{Crypto-Vehicle Trade Example}
    \label{fig:crypto_vehicle_chart}
\end{figure}

Figure \ref{fig:crypto_vehicle_chart} shows a diagram of an identified crypto-vehicle trade.\footnote{I additionally show some examples of matched trades from the data in Table \ref{tab:matched_trades} in the Appendix.} Trade 1 and Trade 2 are observed in the data, whereas the transfer between the cryptocurrency sender and receiver is not. The equal size of these trades in terms of Bitcoin, given that they occur within a short window of each other and are of a relatively uncommon size, allows for their identification as a crypto-vehicle trade. These matched trades are considered ``crypto-vehicle" trades as the purpose of buying Bitcoin in these cases is not to gain exposure to Bitcoin's price volatility (i.e. speculative behavior), rather the user's goal is to use Bitcoin as an intermediate step between exchanging one fiat currency for another, sending money across accounts domestically, or sending money internationally. In fact, users with the intent to use Bitcoin as a financial vehicle should wish to minimize their exposure to Bitcoin's price volatility, another reason that the two transactions representing a crypto-vehicle trade often occur within a small time window. Using this algorithm, and the fact that I observe the user's geolocation on both legs of the crypto-vehicle trade, I can observe money being sent internationally through Bitcoin on the Paxful exchange.

\begin{longtable}{l|ccc}
\caption{
{\large Summary Statistics}
}\label{tab:paxful_summary} \\ 
\toprule
\multicolumn{1}{l}{} & All Trades & Matched & Matched -- Cross Border \\ 
\midrule\addlinespace[2.5pt]
N & $89,668,400$ & $14,982,806$ & $6,568,882$ \\ 
Average Trade Size & $\text{\$}92.67$ & $\text{\$}47.07$ & $\text{\$}46.60$ \\ 
Maximum Trade Size & $\text{\$}143,344.74$ & $\text{\$}47,641.18$ & $\text{\$}18,348$ \\ 
Total Volume & $\text{\$}8.3B$ & $\text{\$}705.3M$ & $\text{\$}306.3M$ \\ 
\bottomrule
\end{longtable}

Table \ref{tab:paxful_summary} shows summary statistics for the Paxful data, before and after matching. Around 17\% of cryptocurrency trades are identified as crypto-vehicle trades. Of these, roughly 44\% are cross-border transactions. These matched trades are on average smaller in value than the average trade in the dataset as a whole–consistent with previous research on remittances which finds that remittances tend to be sent in small and frequent transactions.\footnote{see \cite{yang2011migrant}} This is additionally supported by the fact that, despite making up around 17\% of total transactions, cryptocurrency vehicle trades make up only around 8\% of total volume observed in this time period. I show total cryptocurrency volume for the most crypto-active countries throughout the time period, as well as the breakdown between non-vehicle (speculative) trades versus financial vehicle trades in Appendix Figure \ref{fig:flows_by_country}.\footnote{Due to Nigeria's high level of trading volume on Paxful, I check whether my results are sensitive to the exclusion of Nigeria. My results are similar when I exclude Nigeria from the analysis.}

Finally, after identifying all crypto-vehicle trades in the data via this algorithm, I am able to construct a panel of bilateral cryptocurrency flows using geolocation data from each transaction. Given the granularity of the data in terms of time, any aggregation level is theoretically possible, but I choose to aggregate to the week level in order to reduce volatility and minimize zero observations while retaining some level of granularity in the window I examine in my main analysis.

\section{Empirical Strategy}\label{sect:empirics}

\subsection{Background: Cryptocurrency and Remittances}

Migrant remittances represent a significant global financial flow ($\sim$ \$605 billion to low- and middle-income countries in 2021 alone\footnote{World Bank (2021)}) and a migrant's origin country provides a natural channel for international financial flows. While I do not have information in my data on what cryptocurrency is being used for, I plot the overall relationship between cryptocurrency outflows from the US in 2019 and data on the foreign born population residing in the US from the 2019 American Community Survey in Figure \ref{fig:fb_vs_flows}.

\begin{figure}[!h]
    \centering
    \includegraphics[scale=0.5]{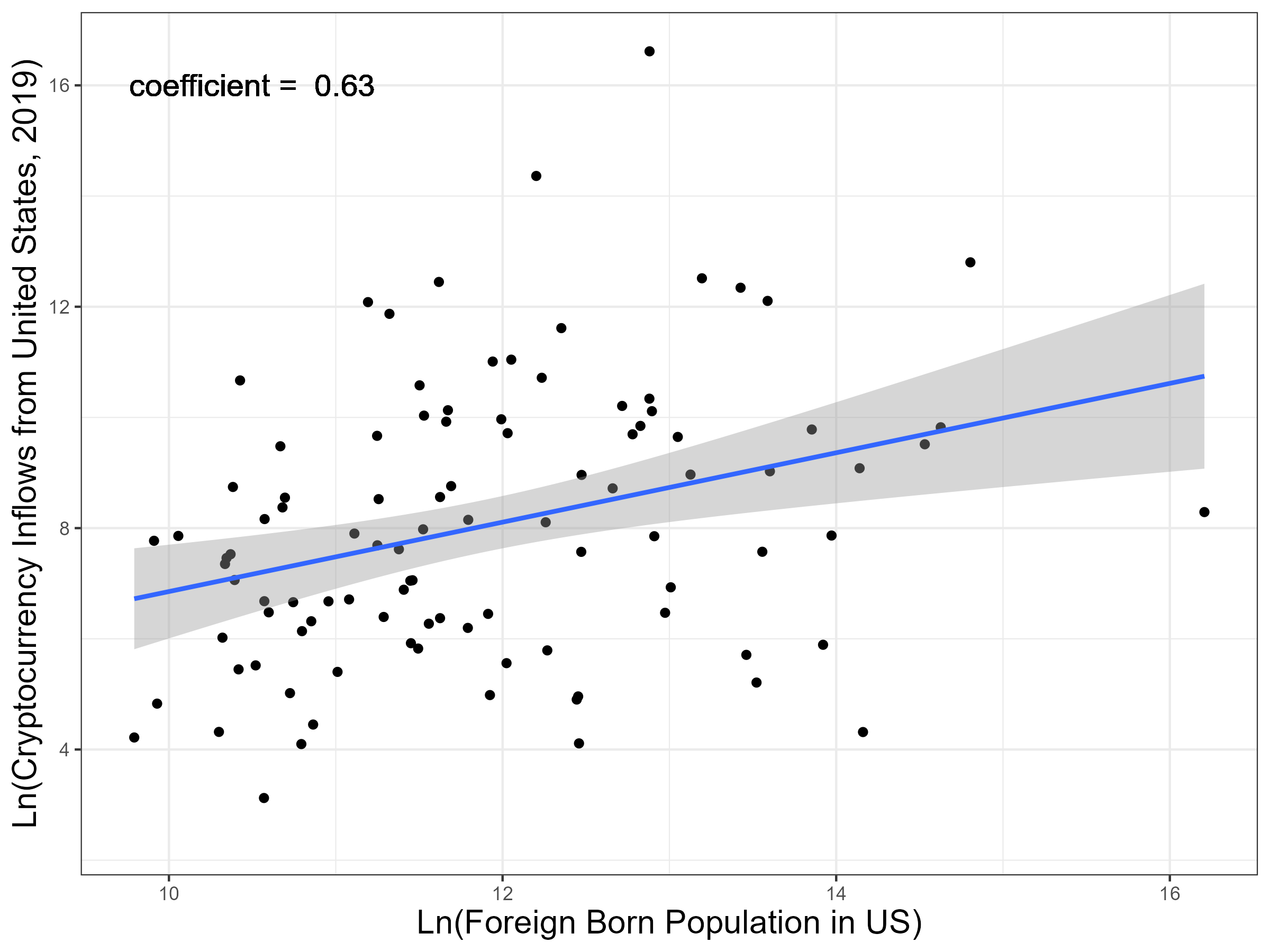}
    \caption{Foreign Born Population vs Cryptocurrency Flows}
    \label{fig:fb_vs_flows}
\end{figure}

This shows a positive association between the foreign-born population and the amount of cryptocurrency received by migrant origin countries, with a 10\% greater foreign-born population associated with 6.3\% higher inflows. This is consistent with research by \cite{freund_remittances_2005} showing that remittances are positively related to the stock of migrants living within a country.

I additionally show the share of the US foreign-born population for the top twenty origin countries in Figure \ref{fig:migrant_shares}, as well as each origin country's income group. The data show that the foreign-born population in the US originates primarily from middle-income countries, making up fifteen out of the top twenty origin countries, whereas high-income origin countries only account for four out of the top twenty, and only one low-income country, Haiti, makes it into the top twenty. This figure suggest middle-income countries are likely candidates for cryptocurrency usage as an international financial vehicle originating from the US, given their significant representation in the US foreign-born population. Additionally, as discussed in the previous section, middle-income countries hit the ``sweet spot" in terms of financial development and technological infrastructure, where users in these countries have enough access to the necessary technology to use cryptocurrency but have an incentive to avoid the formal financial system to avoid fees and delays.

\begin{figure}[!h]
    \centering
    \includegraphics[scale=0.5]{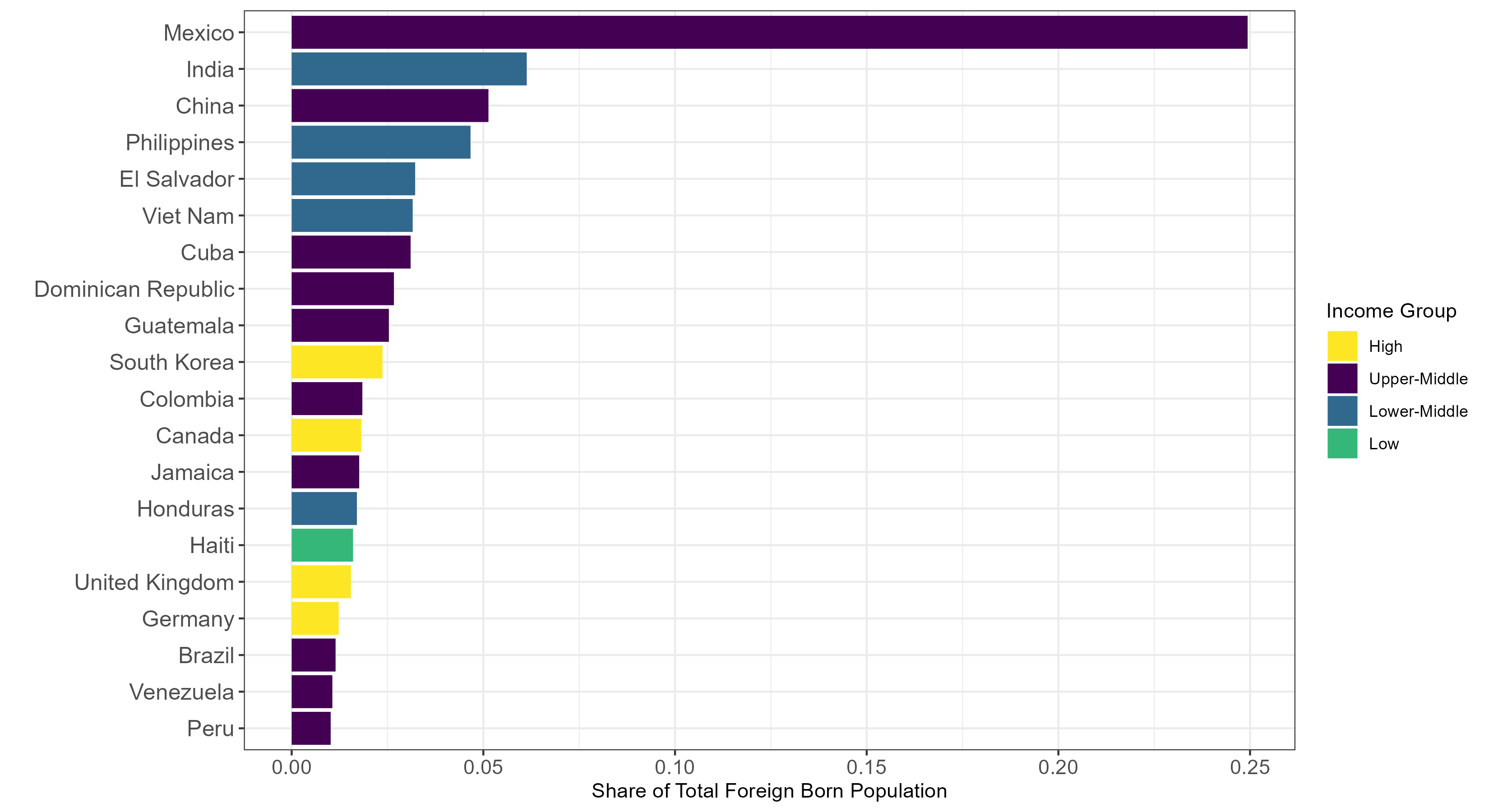}
    \caption{Share of US Foreign Born Population, by Origin Country}
    \label{fig:migrant_shares}
\end{figure}

While I am not making any causal claims with these data, and I do not explicitly tie cryptocurrency flows to remittances in my main analysis, Figures \ref{fig:fb_vs_flows} and \ref{fig:migrant_shares} provide suggestive descriptive evidence that informal remittances are a likely mechanism for fiscal spillovers in this context, and motivate me to investigate heterogeneity by country income group in my main analysis.

\subsection{Background: Stimulus Checks}\label{sect:background}

Due to the onset of COVID-19 and the economic disruptions that followed, the US government issued three rounds of direct fiscal stimulus to the American people via the Coronavirus Aid, Relief, and Economic Security (CARES) Act, which was signed into law on March 27, 2020. There were three rounds of Economic Impact Payments (EIPs or stimulus checks) of varying sizes. The first round of \$1200 stimulus checks were disbursed in April 2020, a second round of \$600 stimulus checks in January 2021, and a third round of \$1400 stimulus checks in March 2021. For the first two rounds of stimulus checks, i.e. April 2020 and December 2020/January 2021, all US citizens, permanent residents, or qualifying resident aliens were eligible to receive the stimulus. To qualify as a resident alien, one needed to have a valid visa, a social security number, and met the substantial presence test (31 days in the US in the current year, and 183 days in the past three years). Undocumented immigrants were thus excluded from receiving stimulus checks, even if they filed taxes. Only US citizens were eligible for stimulus checks in the third round. 

The rollout of stimulus checks was not instantaneous. For the first round of stimulus, payments were sent beginning on April 9, 2020, but many received their payments in the following weeks, or even the following months in some cases. I focus only on the first round of stimulus checks, as later rounds may be confounded by the ongoing impacts of previous rounds of stimulus, as well as other programs rolled out in response to the COVID pandemic.

\begin{figure}[!h]
    \centering
    \includegraphics[scale = 0.6]{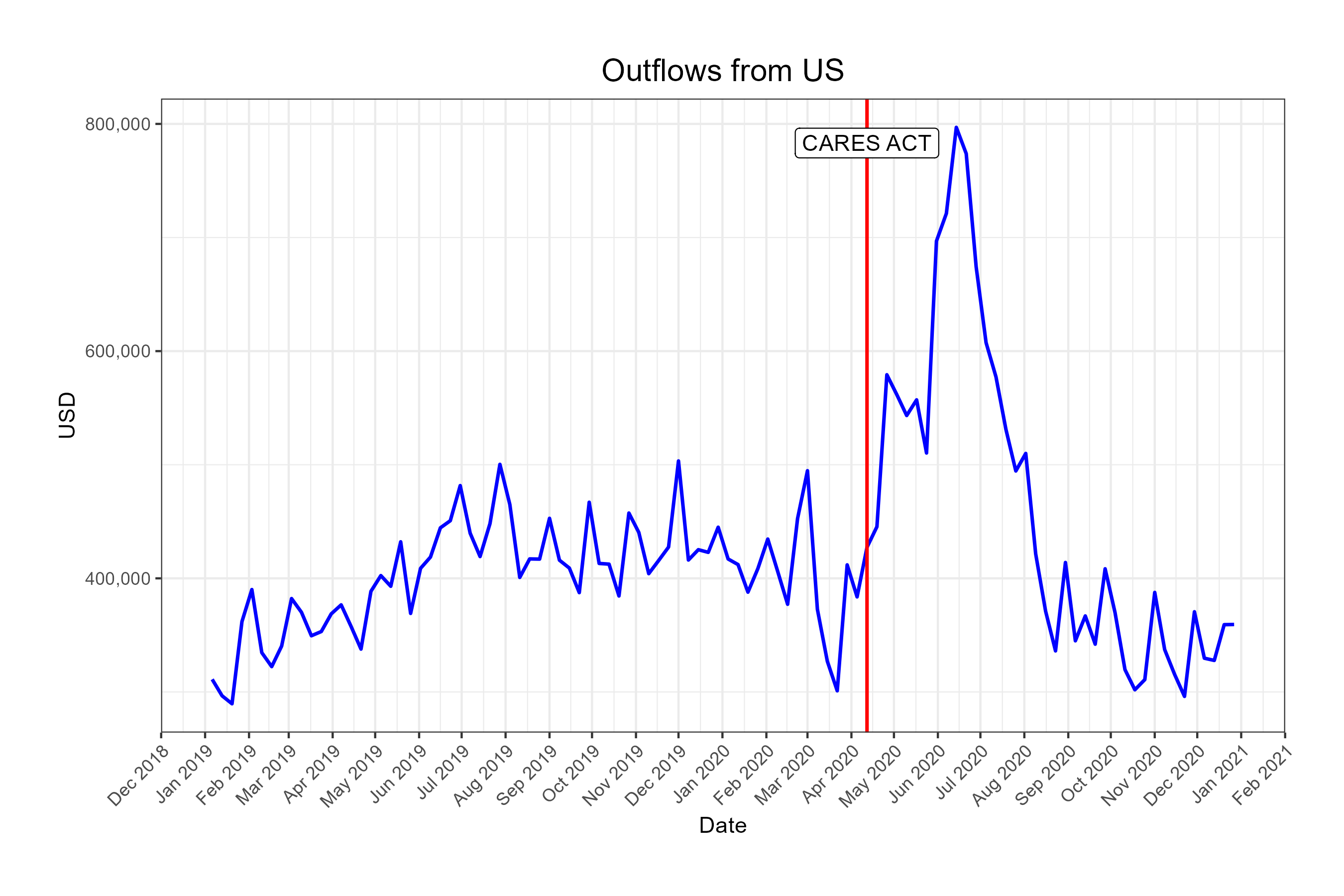}
    \caption{Simulus Timing}
    \floatfoot{Source: Paxful, Author's calculations}
    \label{fig:stimulus_timing}
\end{figure}

As shown in Figure \ref{fig:stimulus_timing}, I observe a temporary and large increase in cryptocurrency outflows in the US following the disbursement of EIPs--with almost double the outflows at its peak compared the period just prior to the stimulus checks. Of course, this could simply reflect global trends in cryptocurrency flows, hence the need for a counterfactual comparison. One may be concerned, given evidence of increased retail trading of Bitcoin due to the stimulus checks found by \cite{divakaruni_uncovering_2021}, that this may simply represent a speculative bubble in Bitcoin trading. First, the increase in speculative retail trading found by \cite{divakaruni_uncovering_2021} is relatively small, and secondly, as previously discussed and shown in Figure \ref{fig:weekly_volume}, the price of Bitcoin is relatively stable during the period of study, limiting the probability that the increase in outflows from the US following the disbursement of EIPs is due to speculative behavior. To rule out speculative behavior as the driver of my results, I include regressions in the Appendix Table \ref{tab:btc_price} in which I include the weekly USD/Bitcoin exchange rate. My main results on the impact of the stimulus checks are unchanged, though interestingly I do find that cryptocurrency outflows are positively related to the USD/BTC price, which could suggest that those who use cryptocurrency as a financial vehicle may also use it as a speculative one. I leave exploring this possibility to future work, as it is beyond the scope of this study.

A potential limitation of this study is that I cannot explicitly disentangle the disbursement of stimulus checks from other programs included in the CARES Act nor variation across U.S. states in terms of their response to COVID, be it the extent of lockdowns to prevent the spread of COVID or additional support to citizens in addition to the CARES Act. With regards to additional programs laid out in the CARES Act, the two provisions I would be most concerned about confounding my analysis are expanded unemployment insurance (UI), which expanded eligibility for unemployment insurance to gig workers, those affected by layoffs, and freelancers in addition to providing an additional \$600 per week in payouts, and the Payment Protection Program (PPP), which provided businesses with loans and grants to help cover their payroll.

While I cannot explicitly rule out that my empirical strategy may capture some of the effects of these additional programs, previous literature studying direct fiscal stimulus suggests that direct, lump-sum stimulus payments have relatively immediate effects on consumption, while policy measures such as UI or payroll protection typically operate with lags.\footnote{see e.g. \cite{sahm2012}, \cite{parker2013}, or \cite{broda2014} for evidence on the speed of impacts from direct vs. indirect fiscal stimulus.} Given the high frequency of my data and the tight window I examine around the disbursement of stimulus checks, this should increase the likelihood that I am capturing the effect of the direct fiscal stimulus of EIPs as opposed to other programs. Additionally, I argue that, between the stimulus checks, expanded UI, and the PPP, the stimulus checks are the most likely to be sent internationally and thus detected by my analysis. While expanded unemployment insurance does represent a positive income shock similar to the stimulus checks, it is limited to a much smaller segment of the population than the stimulus checks and targeted towards jobless individuals, who are much less likely to be in a financial position to remit their stimulus checks. The stimulus checks, on the other hand, were given to all but the highest income individuals and were not targeted depending on an individual's employment status, thus for workers who did not lose their jobs as a result of COVID, the stimulus check represented an instantaneous positive wealth shock that could much more easily be remitted. With regards to PPP loans and grants to businesses, these funds were primarily used to maintain existing payrolls, thus not representing a positive wealth shock to workers in the same way that the stimulus checks did, but rather smoothing workers' income over the negative economic impacts of COVID.

States also varied significantly in terms of their response to COVID, and while I cannot take advantage of this variation due to the fact that cryptocurrency transactions are measured at the country level, it is worth discussing state-level variation in the context of my empirical strategy.\footnote{See \cite{covid_variation_2021} for a systematic analysis of state-level responses to COVID.} Most states that implemented stay-at-home orders implemented them around the end of March 2020 or beginning of April 2020, around the same time as the CARES Act and subsequent disbursement of stimulus checks, with the earliest re-openings occurring around the end of April 2020. To my knowledge, there were no state-level direct stimulus programs implemented during the period I study. If there is an effect of state-level COVID lockdowns on cryptocurrency outflows from the US, I would expect that effect to be negative, which would bias my analysis towards not finding an effect of stimulus checks on cryptocurrency outflows. Moreover, in Section \ref{sect:event_study} I show event-study estimates around the disbursement of stimulus checks and find no clear evidence of a negative impact of state-level lockdowns at the end of March on cryptocurrency outflows nor of a positive impact at the end of April when many states lifted their stay-at-home orders.

\subsection{Difference-in-Differences}\label{sect:did}

In this section I describe the difference-in-differences (DID) framework I use to estimate the effect of the US direct fiscal stimulus on cryptocurrency outflows from the US. This approach compares the trend in cryptocurrency outflows from the US to the trend in cryptocurrency outflows from countries in a comparison group. I use the disbursement of EIPs as the ``treatment" in this case, and observe the response in cryptocurrency outflows from the US relative to the counterfactual. For the DID estimator to produce a valid causal estimate, the assumption of equal counterfactual trends in the absence of treatment must be satisfied. While this assumption is not directly testable, I obtain event-study estimates to observe how similarly the treatment and control group are behaving in the pre-treatment period.

\subsection{Control Group Definition}

Generally, comparisons across countries using the DID methodology are difficult given the difficulty of finding a valid comparison group, especially for a country such as the US. I address this in my main analysis by restricting the comparison group to only be other high-income countries, following the income classifications of the World Bank. This approach constructs a control group that is much more similar to the United States while maintaining a reasonable sample size. The main DID analysis will thus compare the trend in total cryptocurrency outflows from the US to the trend in total cryptocurrency outflows from other high-income countries. Under the assumption of equal counterfactual trends, an increase in outflows from the US relative to outflows from the control group indicates that the disbursement of EIPs caused an increase in cryptocurrency outflows. I show results using only OECD countries as the control group in Section \ref{sect:robustness}.

Ideally, the US would be the only country to have enacted a fiscal stimulus in response to the COVID-19 pandemic. This is obviously not the case, as most governments around the world had some sort of economic response to the crisis. Indeed, three other countries, Japan, Singapore, and South Korea had their own form of direct, one-time stimulus payments occurring in 2020. I remove these countries from the control group. Still, one might be concerned about non-direct stimulus measures such as unemployment insurance or payroll support affecting the ``untreated" status of countries in the control group. I make a similar argument as in Section \ref{sect:background}, that the direct and lump-sum nature of the US COVID stimulus lends itself to a faster response from cryptocurrency senders than other types of stimulus which may take time to work themselves through through the economy down to consumers and would-be cryptocurrency/remittance senders. By using a relatively tight window around the disbursement of EIPs in the US, I hope to mitigate the chance that other, non-direct stimulus affects the control group. Even if it does, the bias should be towards not detecting an effect for US cryptocurrency senders, given that I would expect to see an increase in outflows from other countries as a result of their own fiscal stimulus.

\subsubsection{Estimating Equation}

For my main specification I run a two-way fixed effects (TWFE) regression with an interaction term to capture the treatment effect. I follow \cite{divakaruni_uncovering_2021} in limiting my period of study to January 1, 2020 to June 7, 2020, after which most stimulus checks have been disbursed.

As countries vary significantly in their level of cryptocurrency flows, I opt for Poisson Quasi-Maximum Likelihood Estimation (QMLE) in order to obtain a percent-change interpretation.\footnote{after transforming estimates via $e^{\beta} - 1$.} I use Poisson QMLE to account for zeros in my data, as there are many country-by-time observations that record zero flows. As zeros are not defined when using a log transformation, and recent work has shown that log-like transfomations such as $ln(Y+1)$ or $arcsinh(Y)$ are not scale-invariant, or in other words that estimated coefficients can be scaled arbitrarily by changing the units in which outcomes are measured, I opt for Poisson QMLE instead.\footnote{See \cite{chen2023logs} for an in-depth discussion of log-like transformations, their limitations, and alternatives.} I also need not worry about issues brought to light in the recent TWFE literature, as in this case there is only one treated unit receiving treatment at a single time, thus there is not staggered adoption to account for.\footnote{See \cite{BAKER2022370} for an explanation on issues with staggered difference-in-differences designs using two-way fixed effects.}

I estimate the following via Poisson QMLE:

\begin{align}\label{eq:qmle_did}
    CryptoOutflows_{it} = \exp(\beta \left[Disbursed_{t} \times US_{i}\right] + \alpha_{i} + \alpha_{t})\epsilon_{it}
\end{align}



where the dependent variable, $CryptoOutflows_{it}$ represents the total volume, measured in USD, of cryptocurrency outflows from country $i$ in week $t$. $Disbursed$ takes on a value of one after the disbursement of EIPs on April 9, 2020, and zero beforehand. These variables are interacted with $US_{i}$, which takes on a value of one if country $i$ is the United States and zero otherwise. Country and time period fixed effects are represented by $\alpha_{i}$ and $\alpha_{t}$, respectively, and absorb time-invariant country characteristics and time trends in cryptocurrency outflows common to all countries. Finally, $\epsilon$ represents the idiosyncratic error term. The main parameter of interest is thus $\beta$, the difference-in-differences estimator. Note that this functional form gives an estimate of $\beta$ that is in terms of a difference in logs, so in order to achieve a relative percent change interpretation one must transform $\beta$ via $\theta_{ATE\%} = e^{\beta} - 1$.

Given the discussion about ``crypto suitability" in Section \ref{sect:data_paxful}, I investigate whether the response of cryptocurrency outflows to the disbursement of EIPs depends on where outflows are destined, and particularly on the development level of the destination. To do this, I construct several measures of $CryptoOutflows_{it}$. The base case is to construct cryptocurrency outflows with no restriction on the destination, where I do not eliminate any cross-border trades prior to aggregating. I additionally construct cryptocurrency outflows destined to high-income countries, middle-income countries, and low-income countries by respectively restricting the aggregation to trades where the receiving party is located in a country in the respective category. For example, to construct outflows to high-income countries, I aggregate only trades where the receiving user is located in a country classified as high-income by the World Bank.

For this regression to give me the causal effect of the stimulus checks on cryptocurrency outflows, I need to assume that the comparison group, i.e. the non-US countries in my sample, represent a valid counterfactual to the US, or in other words, that the parallel trends assumption is satisfied. Intuitively, this assumes that average percentage change in the US's cryptocurrency outflows would have been similar to the average percentage change in cryptocurrency outflows of the comparison countries, had it not been for the disbursement of stimulus checks. While this assumption in not explictly testable, I plot event-study estimates in Section \ref{sect:event_study} by interacting $US_{i}$ with weekly dummy variables to obtain separate DID estimates for each week. This allows me to observe how the US's cryptocurrency outflows are evolving relative to the comparison group in the time leading up to the disbursement of stimulus checks, and to observe the dynamics of US cryptocurrency outflows over time in response to the stimulus checks. For example, I use these event-study estimates to investigate how persistent any increase in outflows due to the stimulus checks is.

\section{Results}\label{sect:results}

In this section I present my main results. I find evidence that cryptocurrency outflows from the US increased significantly relative to the control group following the disbursement of stimulus checks, particularly for cryptocurrency flows destined to middle- and high-income countries. I find no evidence of an effect for cryptocurrency flows destined to low income countries. Event-study estimates provide evidence that the parallel trends assumption is satisfied. Time dynamics suggest that the increase in cryptocurrency outflows is temporary.

\subsection{DID Estimates}\label{sect:did_results}

\begin{table}[!h]
\centering
\begin{threeparttable}
\caption{Poisson QMLE–Dependent Variable: Cryptocurrency Outflows}\label{tab:twfe_qmle}
\centering
\begin{tabular}[t]{lcccc}
\toprule
  & All Destinations & Low-Income & Middle-Income & High-Income\\
\midrule
\addlinespace[0.5em]
\multicolumn{5}{l}{\textit{Control Group: High-Income Countries}}\\
\midrule \hspace{1em} $\hat{\theta}_{ATE\%} = e^{\hat{\beta}} - 1$ & 0.114** & -0.072 & 0.177*** & 0.158***\\
\hspace{1em} & (0.047) & (0.242) & (0.054) & (0.052)\\
\hspace{1em}$\text{Observations}$ & 1357 & 621 & 1265 & 1265\\
\hspace{1em}Country FE & X & X & X & \vphantom{1} X\\
\hspace{1em}Week FE & X & X & X & \vphantom{1} X\\
\addlinespace[0.5em]
\bottomrule
\multicolumn{5}{l}{\rule{0pt}{1em}* p $<$ 0.1, ** p $<$ 0.05, *** p $<$ 0.01}\\
\end{tabular}
\begin{tablenotes}
\small
\item [a] Standard errors clustered at the country level.
\end{tablenotes}
\end{threeparttable}
\end{table}

Table \ref{tab:twfe_qmle} shows the results from running Equation \ref{eq:qmle_did}, using the period from January 1, 2020 to June 7, 2020. Standard errors are clustered at the country level. Going across columns, each column shows the results using varying aggregation schemes for constructing cryptocurrency outflows. The column ``All Destinations" uses cryptocurrency outflows aggregated to the country-week level with no restrictions on the receiving destination of cryptocurrency transactions. The column ``Low-Income" performs the same aggregation but only for cryptocurrency transactions for which the receiving country is classified as low income by the World Bank. The same logic applies to the ``Middle-Income" and ``High-Income" columns. In all columns, the comparison is between the United States as the treated unit, and other high-income countries as the control group. Putting it all together, and using the ``Middle-Income" column as an example, this regression compares the trend in outflows originating from the US and arriving in middle-income countries to outflows originating from other high-income countries and arriving in middle-income countries, before and after the disbursement of EIPs.

I present coefficients in terms of their percentage change interpretation by transforming estimates via $\hat{\theta}_{ATE\%}=e^{\hat{\beta}} - 1$. To interpret these coefficients, I again take the ``Middle-Income" column as an example. This regression gives an estimate of $e^{0.163} - 1 \approx 17.7\%$ and is significant at the one percent level. The interpretation here is that, cryptocurrency outflows destined to middle-income countries increased by 17.7\% more from the US than cryptocurrency outflows destined to middle-income countries originating from high-income countries, after the disbursement of EIPs. Standard errors are transformed via the delta method.

My results suggest that cryptocurrency outflows from the US increased significantly as a result of the disbursement of EIPs. I find an overall increase of cryptocurrency outflows to all destinations of 11.4\%, as shown in the first column. In the second column, where cryptocurrency outflows are limited to those destined for low-income countries, there is no evidence of an increase caused by the disbursement of EIPs, as the coefficient is negative but statistically insignificant from zero. The third column shows a large increase of 17.7\%, significant at the one percent level, for cryptocurrency outflows from the US destined for middle-income countries. The last column shows a slightly smaller increase of 15.8\%, statistically significant at the one percent level, for cryptocurrency outflows from the US destined for high-income countries. Taken together, my results suggest that the disbursement of EIPs in the US is associated with a significant increase in cryptocurrency outflows from the US in the subsequent time period, particularly for cryptocurrency destined to middle- and high-income countries.

One thing to note is that the number of observations changes when moving across columns, even though the number of time periods and theoretical control units does not. The reason for the varying number of observations is that, if for all time periods a country in the comparison group records no outflows at all, then that country drops out of the analysis due to the inclusion of country fixed effects. The fact that the number of observations for the ``Low-Income" regressions are roughly half that of the other categories indicates that there are a significant number of countries in the control group that record zero flows to low-income destinations throughout the entire period of study.

\subsection{Event-Study Estimates}\label{sect:event_study}

\begin{figure}[h!]
    \centering
    \includegraphics[width=0.8\textwidth]{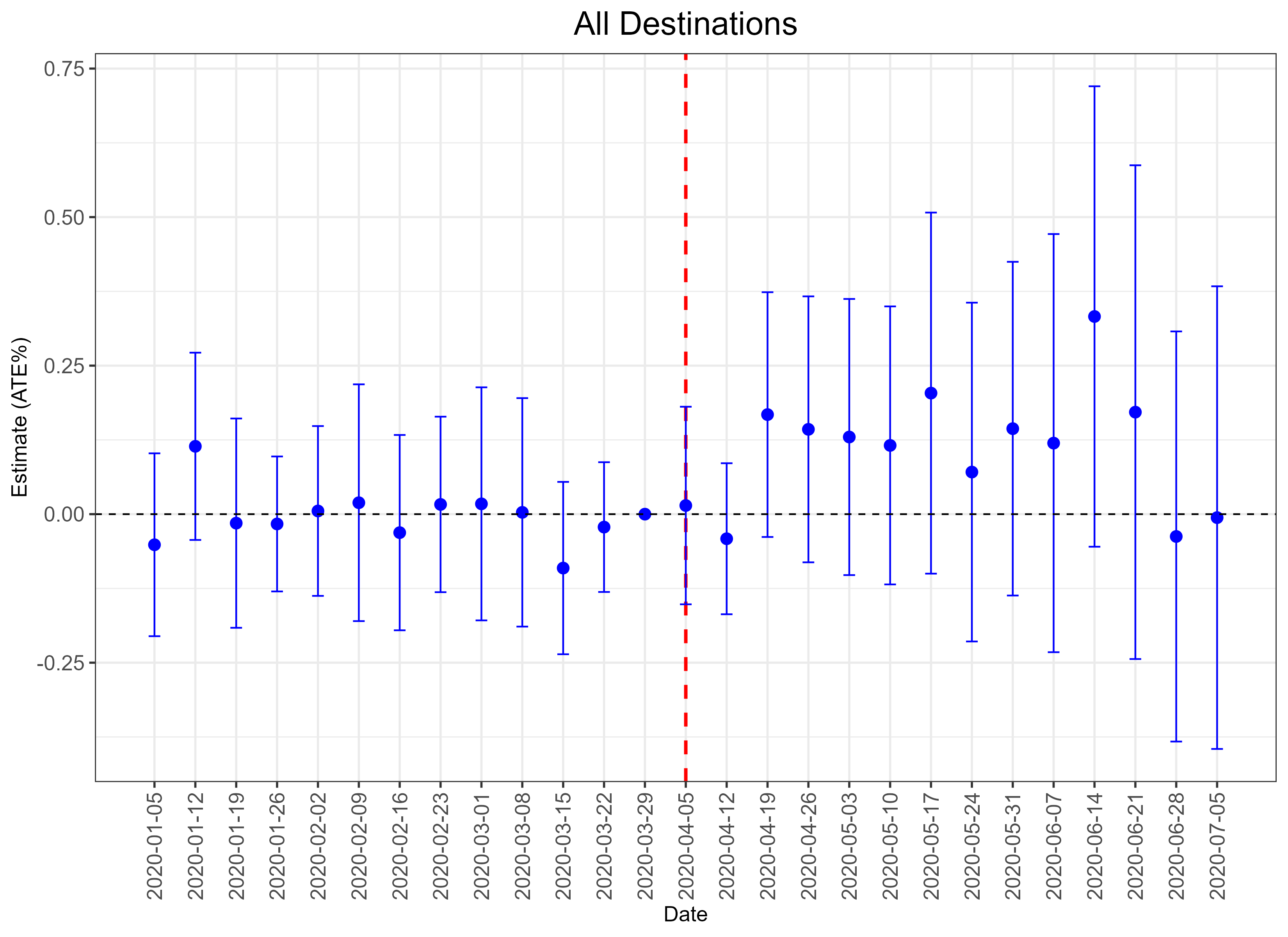}
    \caption{Event-Study Estimates}
    \label{fig:es_plot_all_destinations}
    \floatfoot{Figure \ref{fig:es_plot_all_destinations} shows event-study estimates for Equation \ref{eq:qmle_did}. Estimates are obtained by interacting a dummy for each week with a dummy variable that takes on a value of one for the US and zero otherwise, as well as week and country fixed effects. I use the week prior to the disbursement of stimulus checks as the reference period. The vertical dotted line corresponds to the week during which stimulus checks began being disbursed. Dates on the horizontal axis indicate the first day of each week of aggregation, so `2020-01-05' refers to the week starting with January 5th, 2020, and so on. Error bars show 95\% confidence intervals.} 
\end{figure}

To test for differential pre-treatment trends between the US and the control group, and to observe the dynamics of treatment effects over time, I show event-study estimates in Figure \ref{fig:es_plot_all_destinations}. As in Table \ref{tab:twfe_qmle}, I transform estimates into their percent-change interpretation using $e^{\beta} - 1$, and use the delta method to transform standard errors. This event-study uses cryptocurrency outflows with no restriction on destination, corresponding to the regression shown in the first column of Table \ref{tab:twfe_qmle}. I use the week immediately prior to the disbursement of stimulus checks as my reference period, thus each event-study estimate shows the difference in cryptocurrency outflows, in percentage terms, between the US and the control group in a given week relative to the gap between the US and control group in the week prior to disbursement. Estimates in the pre-treatment period provide evidence that the high-income countries represent a good counterfactual to the US, with most point estimates being very close to zero.

After the disbursement of stimulus checks, represented by the vertical dotted line, the point estimates show a sharp increase in cryptocurrency outflows relative to the control group, with estimates hovering around a 20-25\% increase in weekly cryptocurrency outflows for the two months after EIPs are disbursed. The increase in outflows appears to be temporary, as estimates return to zero around the end of June 2020. While the 95\% error bars do cover zero for these estimates, it should be noted that in a given week there is a limited number of observations, which reduces precision. In the main estimates in Section \ref{sect:did_results}, a statistically significant increase in outflows due to the stimulus checks is found.

\begin{figure}[h!]
    \centering
    \begin{subfigure}{0.49\textwidth}
        \includegraphics[width = \textwidth]{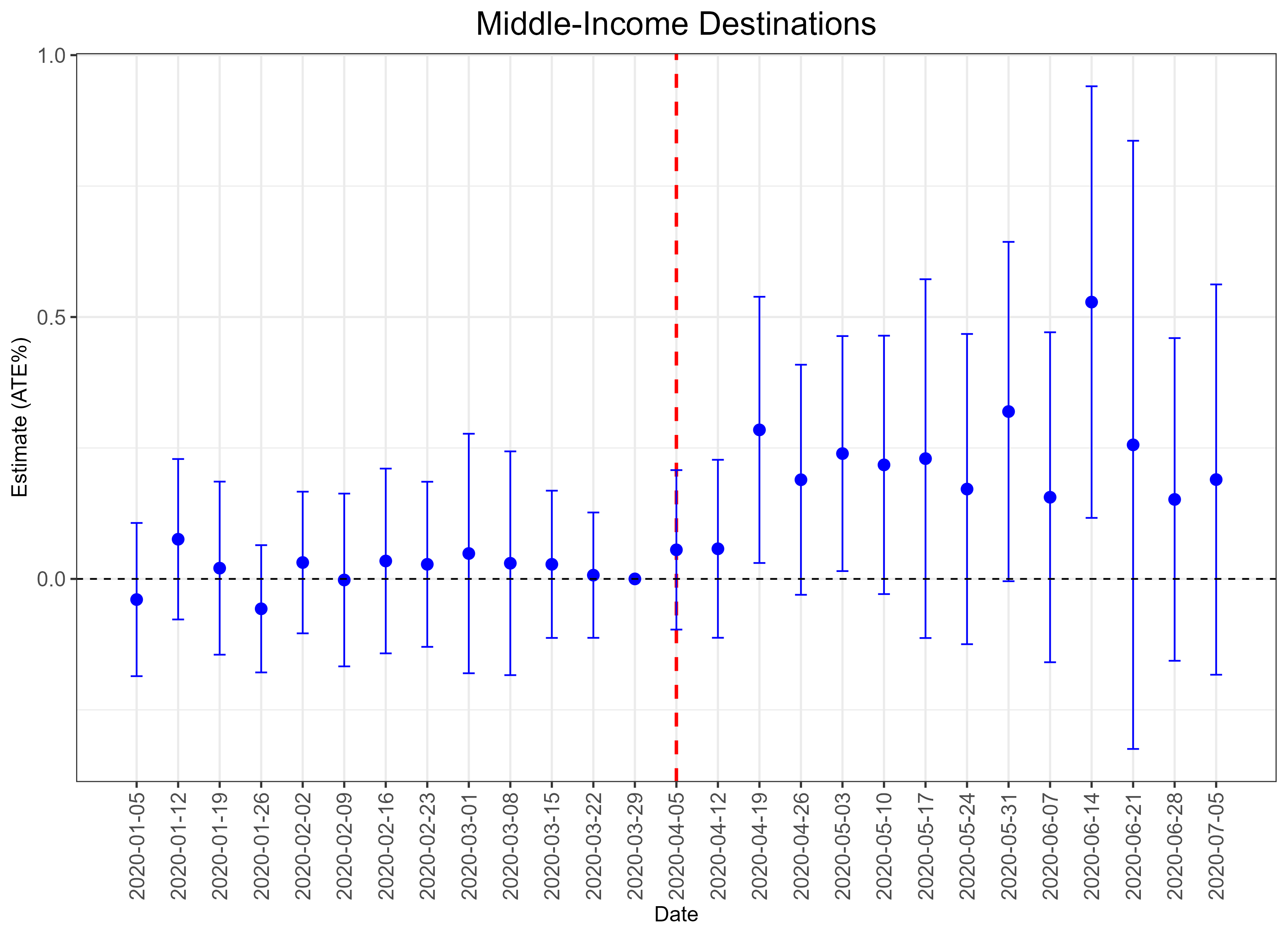}
    \end{subfigure}
    \begin{subfigure}{0.49\textwidth}
        \includegraphics[width = \textwidth]{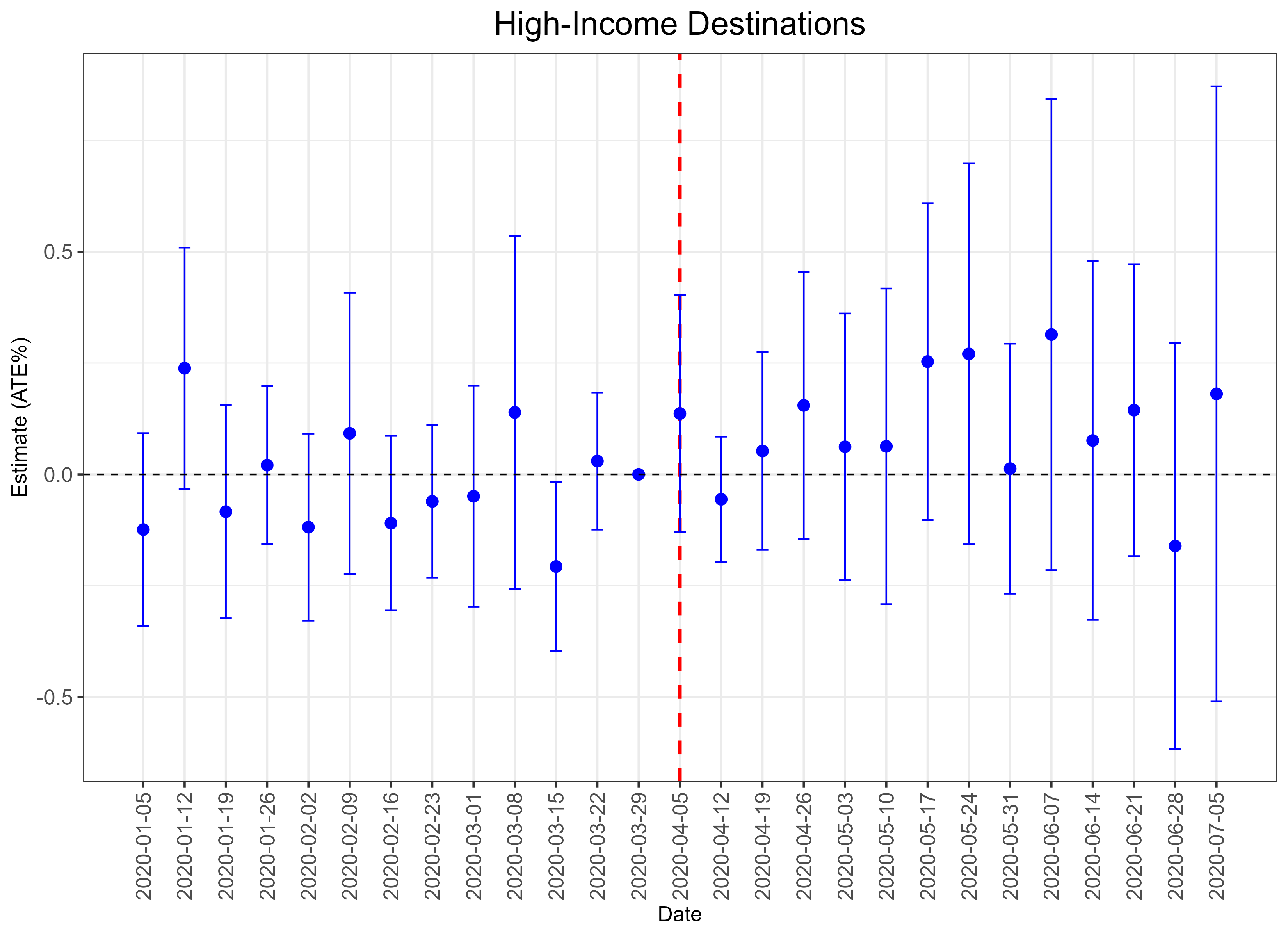}
    \end{subfigure}
    \caption{Event-Studies By Destination Income Level}\label{fig:es_plots_by_income}
    \floatfoot{Figure \ref{fig:es_plots_by_income} shows event-study estimates for Equation \ref{eq:qmle_did}. Estimates are obtained by interacting a dummy for each week with a dummy variable that takes on a value of one for the US and zero otherwise, as well as week and country fixed effects. I use the week prior to the disbursement of stimulus checks as the reference period. The vertical dotted line corresponds to the week during which stimulus checks began being disbursed. Dates on the horizontal axis indicate the first day of each week of aggregation, so `2020-01-05' refers to the week starting with January 5th, 2020, and so on. Error bars show 95\% confidence intervals.} 
\end{figure}

In Figure \ref{fig:es_plots_by_income} I show event-study estimates separately for cryptocurrency outflows destined to middle- and high-income countries, corresponding to the third and fourth columns in Table \ref{tab:twfe_qmle}. The left panel, which shows estimates for outflows destined to middle-income countries, tells a similar story to Figure \ref{fig:es_plot_all_destinations}. Estimates in the pre-period are very close to zero, indicating that the trajectories of cryptocurrency outflows for the US follows that of the control group closely in the time periods leading up to treatment. After the disbursement of EIPs, a sharp increase in cryptocurrency outflows from the US can be seen, with estimates again decreasing near the end of June 2020. The increase is slightly larger than that seen in Figure \ref{fig:es_plot_all_destinations}, with estimates floating around 25\% for the two months following the disbursement of EIPs. This is consistent with the results shown in Table \ref{tab:twfe_qmle}, where the largest increase in cryptocurrency outflows is seen for those destined to middle-income countries.

The right panel of Figure \ref{fig:es_plots_by_income}, which shows event-study estimates for cryptocurrency destined to high-income countries, displays a larger variation in point estimates, but ultimately a similar story. Point estimates bounce around zero in the pre-period, and are generally positive immediately after the start of treatment, with estimates returning to zero near the end of June 2020. Finally, consistent with the second column of Table \ref{tab:twfe_qmle}, estimates for cryptocurrency outflows destined to low-income countries are measured with considerable noise due to the low amount of observations and are thus uninformative, but are made available in the Appendix.

\section{Discussion}\label{sect:discussion}

Taken as a whole, my results suggest that the disbursement of direct stimulus payments as appropriated by the CARES Act is associated with a sharp but temporary increase in cryptocurrency outflows, providing evidence of fiscal spillovers through this financial channel. Thus far, however, my analysis has only examined cryptocurrency flows in a tight window around the disbursement of stimulus checks, in order to increase the likelihood of causally identifying their impact on cryptocurrency outflows. In this section, I perform an exploratory exercise where I attempt to quantify the overall size of fiscal spillovers through this channel associated with the CARES Act. As this requires making assumptions about the representativeness of Paxful to the broader cryptocurrency market or to remittance behavior more broadly, that I am unable to verify due to data availability, I urge the reader to interpret this analysis with caution. Rather, the goal of this exercise is to show the \textit{potential} size of spillovers through this channel under various scenarios and assumptions.

\subsection{Quantifying Fiscal Spillovers}

I start by quantifying the overall increase in cryptocurrency outflows from the US due to the disbursement of EIPs for the entire year of 2020. The reasoning behind this is that macro data at the country level is often measured at the annual level, thus getting an estimate for the entire year will allow me to make comparisons with other available data. To do this, I extend the event-study shown in Section \ref{sect:event_study} to cover the entire year of 2020, using cryptocurrency outflows to all destinations as the outcome. I use these event-study estimates to construct counterfactual cryptocurrency outflows for the US for the entire year. I estimate the overall increase in cryptocurrency outflows in 2020, by summing together, for each week, the difference between observed US cryptocurrency outflows and counterfactual cryptocurrency outflows, then dividing by the sum of observed outflows:

\begin{equation}
    \text{Total Spillover}(\%) = 100 \times \frac{ \sum_{t = 1}^{T} \left[Y_{t}(1) - \hat{Y}_{t}(0)\right]}{ \sum_{t=1}^{T} Y_{t}(1)} \approx 1.27\%
\end{equation}

where $Y_{t}(1)$ represents observed cryptocurrency outflows from the US for week $t \in \{\text{2020-01-01,...,2020-12-31}\}$, and $\hat{Y}_{t}(0)$ represents the untreated counterfactual cryptocurrency outflows for the US in week $t$. This calculation gives an estimate of around 1.27\%, suggesting that around 1.27\% of total US cryptocurrency outflows on Paxful in 2020 are associated with the disbursement of EIPs. I next attempt to generalize these findings to get an overall estimate of the size of fiscal spillovers through this channel relative to the spending on EIPs allocated by the CARES Act.

\begin{table}[!h]
\centering
\resizebox{\textwidth}{!}{%
\begin{tabular}{@{}ccccc@{}}
\toprule
 &
  2020 Total Volume &
  \begin{tabular}[c]{@{}c@{}}Estimated \\ 2020 US Outflows\end{tabular} &
  \begin{tabular}[c]{@{}c@{}}Estimated \\ Fiscal Spillover\end{tabular} &
  \begin{tabular}[c]{@{}c@{}}Fiscal Spillover, Relative to \\ \$271.4B CARES Act EIPs\end{tabular} \\ \midrule
\multicolumn{1}{c|}{Paxful}                         & \$1.9B   & \$23.0M  & \$292,798 & Negligible \\
\multicolumn{1}{c|}{Top 15 Crypto Exchanges}        & \$3.78T  & \$45.8B  & \$583.0M  & 0.215\%    \\
\multicolumn{1}{c|}{Total Cryptocurrency Market}    & \$44.42T & \$537.7B & \$6.84B   & 2.52\%    \\
\multicolumn{1}{c|}{Official Remittances (World Bank)} &          & \$66.54B & \$847.1M  & 0.312\%    \\
\multicolumn{1}{c|}{Informal Remittances (Lower Bound)} &          & \$33.27B & \$423.5M  & 0.156\%    \\
\multicolumn{1}{c|}{Informal Remittances (Upper Bound)} &          & \$166.35B & \$2.12B  & 0.78\%   
\end{tabular}%
}
\caption{Fiscal Spillovers}\label{tab:fiscal_spillovers}
\end{table}

Table \ref{tab:fiscal_spillovers} shows back-of-the-envelope calculations of fiscal spillovers under varying assumptions. For Paxful itself, the total volume of transactions, measured in USD, was around \$1.9 billion in 2020. If we limit transaction to only those that represent outflows from the US, about \$23 million, or 1.21\% of Paxful's total volume represents US outflows. If we take only the top 15 cryptocurrency exchanges, as measured by their 2020 volume, and assume that a similar proportion of the volume on these exchanges is used to transfer funds out of the US, this would give an overall US outflow through cryptocurrency of \$45.8 billion.\footnote{Estimates of cryptocurrency volume for the top 15 exchanges come from CoinGecko.}\footnote{This is not an entirely unreasonable assumption. The top cryptocurrency exchanges often have user-friendly interfaces similar to Paxful's which would make it easy for individuals who are not technically adept to use the platform. It is also very easy to send money from one location to another, as all one would need is someone in the receiving location to have a digital wallet. As mentioned previously, these transactions are very difficult to track and regulate.} If we further assume a similar impact from the disbursement of EIPs as that found on Paxful, this would give an overall spillover of \$45.8 billion $\times$ 1.27\% $=$ \$583.0 million. Relative to the \$271.4 billion allocated towards EIPs in the CARES Act, this is a 0.215\% fiscal spillover. If we further assume that the response to EIPs found in this study are true of the entire cryptocurrency market, which had an overall volume of \$44.42 trillion in 2020, the total US outflows through cryptocurrency would come to \$537.7 billion.\footnote{The \$44.42 trillion estimate is according to \href{https://coincodex.com/trading-volume/}{CoinCodex}} This figure would give an overall fiscal spillover of \$537.7 billion $\times$ 1.27\% $=$ \$6.84 billion, or 2.52\% of US spending on EIPs from the CARES Act.

Rather than using cryptocurrency as the baseline to calculate fiscal spillovers, let us instead assume that the response to the disbursement of EIPs found in this study is true of remittances overall. Then, using the official (formal) outward remittances from the US as reported by the World Bank of \$66.54 billion, the estimated fiscal spillover would come to \$66.54 billion $\times$ 1.27\% $=$ \$847.1 million, or 0.312\% of spending on EIPs in the CARES Act. I show the same calculation for the upper and lower bound of the rough estimates that exist for the size of informal remittances relative to formal remittances. In the lower-bound case, where informal remittances are 50 percent of formal remittances, this comes out to a fiscal spillover that is 0.156\% of total EIP spending. For the upper bound, assuming that informal remittances are 250 percent of formal remittances, this produces an estimated spillover of 0.78\% of EIP spending.

Given that the ``Total Cryptocurrency Market" scenario assumes that US outflows through cryptocurrency are over 8 times that of the formal remittances reported by the World Bank for 2020, far exceeding the upper bound estimate of informal remittances given by the literature, I view this as an absolute upper bound estimate for fiscal spillovers in this context. Given the results implied by the remittance market and the top 15 cryptocurrency exchanges, a more realistic estimate for the size of spillovers is likely within the 0.2\%-0.7\% range.

\subsection{Implications for Policy}

Taken as a whole, my results suggest that fiscal spillovers, while present, are likely not substantial in size relative to total expenditure. For policymakers in the US responding to a crisis such as COVID-19, where drastic action is needed to bolster the economy, other concerns such as identity fraud, speed of disbursement, or the targeting of aid towards households most in need may be a higher priority. Even in the presence of large fiscal multipliers, my findings suggest that fiscal spillovers likely decrease the fiscal base by less than one percent, thus the vast majority of fiscal expenditure participates in the virtuous cycle of spending that ultimately boosts the economy.

As a general matter, it is worth noting that this study provides evidence that digital currencies are being used for the sending and receiving of remittances. This suggests that the lower fees, timely settlement, and always-available nature of digital currencies make this an attractive alternative to traditional remittance channels. Countries looking to encourage inward remittances can adopt policies to make digital currencies easier to use and access, by reducing barriers for cryptocurrency to fiat conversions or by supporting digital infrastructure such as internet access.

The tax implications of remittances flowing through cryptocurrency as opposed to traditional remittance channels depend on the tax treatment of international transfers in a particular country. Countries that typically tax international money transfers may wish to discourage remittances through cryptocurrency if a the implied tax loss is large enough. For the US, outward remittances are generally not subject to a tax, indeed transactions less than \$10,000 are typically not reported to the IRS. Given that, on Paxful and in general, remittances typically consists of small, frequent transfers, there would not be a large tax impact if remittances begin using cryptocurrency as a financial channel. In a world where all remittances are sent via cryptocurrency, the tax loss to the US government would mostly be in the form of lost taxes from the revenues of money operators such as Western Union or Wise, although this would be a negligible loss compared to the overall tax base in the US and could be offset in part by tax revenue from cryptocurrency exchanges based in the US.

Finally, while the fiscal spillovers found in this study are small relative the expenditure for the US, they may be sizable for smaller, remittance-receiving countries. The implications for policymakers in these countries depend on policy objectives and current economic conditions. If, as many countries were during the COVID-19 pandemic, an economy is being affected by a negative shock, then these spillovers may be a welcome relief, and allow an economy to ``free ride" off of fiscal stimulus from larger economies. On the other hand, for an economy dealing with a crisis such as hyperinflation, a sudden, positive fiscal shock may be precisely the opposite of what policymakers want, and thus trying to mitigate exposure to other countries' fiscal actions through this channel may be warranted.

\section{Robustness Checks}\label{sect:robustness}

\subsection{OECD Countries as Control Group}

\begin{table}[!h]
\centering
\begin{threeparttable}
\caption{Poisson QMLE–Dependent Variable: Cryptocurrency Outflows}\label{tab:twfe_qmle_oecd}
\centering
\begin{tabular}[t]{lcccc}
\toprule
  & All Destinations & Low-Income & Middle-Income & High-Income\\
\midrule
\addlinespace[0.5em]
\multicolumn{5}{l}{\textit{Control Group: OECD Countries}}\\
\midrule \hspace{1em} $\hat{\theta}_{ATE\%} = e^{\hat{\beta}} - 1$ & 0.118* & 0.007 & 0.161** & 0.175***\\
\hspace{1em} & (0.061) & (0.302) & (0.069) & (0.072)\\
\hspace{1em}$\text{Observations}$ & 805 & 552 & 805 & 805\\
\hspace{1em}Country FE & X & X & X & \vphantom{1} X\\
\hspace{1em}Week FE & X & X & X & \vphantom{1} X\\
\addlinespace[0.5em]
\bottomrule
\multicolumn{5}{l}{\rule{0pt}{1em}* p $<$ 0.1, ** p $<$ 0.05, *** p $<$ 0.01}\\
\end{tabular}
\begin{tablenotes}
\small
\item [a] Standard errors clustered at the country level.
\end{tablenotes}
\end{threeparttable}
\end{table}

In this section, I demonstrate robustness of my main results to a different control group by running Equation \ref{eq:qmle_did} using only OECD countries as a control group. As with the main results, I produce estimates using varying aggregation schemes for the outcome variable, cryptocurrency outflows. Specifically, for each OECD country including the US I aggregate cryptocurrency trades into weekly outflows, first with no restriction on the destination, then restricting the destination of cryptocurrency trades to low-income, middle-income, and high-income countries, respectively. The results are very similar to those found in Section \ref{sect:did_results}, with cryptocurrency outflows from the US increasing by 11.8\% relative to the OECD countries for outflows with no restriction on destination. No effect is found for cryptocurrency destined to low-income countries, and similar effects, in the 16-17\% range, are found for cryptocurrency destined to middle- and high-income countries. Estimates are slightly less precise than those found in Section \ref{sect:did_results} due to the smaller sample size.

\subsection{Alternative Methodology: Synthetic Difference-in-Differences}

In this section I add robustness to my main results using a Synthetic Difference-in-Differences (SDID) methodology, following the seminal paper by \cite{arkhangelsky2021}. SDID combines features of both the synthetic control (SC) estimator as pioneered by \cite{abadie2003} and the traditional DID estimator via the construction of a counterfactual via a weighting scheme.  Specifically, SDID solves:

\begin{align}
    \left(\hat{\tau}^{sdid}, \hat{\mu}, \hat{\alpha}, \hat{\beta}\right) = \argmin_{\tau, \mu, \alpha, \beta} \left\{\sum^{N}_{i=1}\sum^{T}_{t=1}\left(Y_{it} - \mu - \alpha_{i} - \beta_{t} - W_{it}\tau\right)^{2}\hat{\omega}_{i}^{sdid}\hat{\lambda}_{t}^{sdid}\right\}
\end{align}

Where $\hat{\tau}^{sdid}$ represents the SDID treatment effect estimator, $\mu$ represents the intercept, $\alpha_{i}$ is a unit fixed effect, $\beta_{t}$ is a time fixed effect, $W_{it} \in \{0,1\}$ is a binary treatment indicator that takes on a value of one for the US after stimulus checks are disbursed, $\hat{\omega}^{sdid}_{i}$ are control unit weights that align pre-treatment trends between the US and control group, and $\hat{\lambda}^{sdid}_{t}$ are time weights that match pre-treatment time periods with post-treatment periods for the control group.

Unlike the SC estimator, SDID allows for an intercept when constructing counterfactual weights via the inclusion of a unit fixed effect. For this reason SDID is preferable to SC in my setting, as countries vary significantly in their baseline level of cryptocurrency usage, which SDID allows for. Since SDID has no equivalent to Poisson QMLE and I cannot use a log transformation given the zeros in my data, the SDID estimate will be in terms of the level of cryptocurrency volume, measured in USD, rather than a percent change.

SDID allows the inclusion of time-varying covariates in order to improve the fit between the treated unit and synthetic control. A challenge in this setting, then, is that coviariates must vary at the weekly level, and most country-level data on factors relevant to cryptocurrency outflows are typically measured at an annual cadence. This makes the inclusion of covariates such as GDP per capita, cryptocurrency adoption, or measures of internet infrastructure generally infeasible to include. As an alternative, I construct covariates intended capture the \textit{pattern} of outflows in terms of their destination, in addition to the overall pattern of outflows. To do this, I take the top ten destinations for cryptocurrency outflows originating from the US and construct, for each week, the share of US cryptocurrency outflows from the US that arrive there. For each country in the donor group, I calculate the same share for the same set of destination countries. Thus the procedure attempts to construct a synthetic control that not only mathces the US on the pattern of overall outflows, but also where those outflows are destined to.

\subsubsection{Variance Estimation via Placebo Method}

To perform inference, I use the placebo method as described by Algorithm 4 in \cite{arkhangelsky2021}, whereby control units are sent through the procedure as if they were treated, and a distribution of treatment effects is formed from which the variance can be calculated.\footnote{I use the default of 200 placebo repetitions in the ``sdid" R package.} I perform this variance estimation both for the overall SDID estimate, as well as to generate confidence intervals when estimating the dynamics of the effect over time.

\subsubsection{Results}

\begin{table}[!h]
\centering
\resizebox{0.4\textwidth}{!}{%
\begin{tabular}{@{}lcc@{}}
\toprule
\multicolumn{1}{c}{} & (1)      & (2) \\ \midrule
\begin{tabular}[c]{@{}l@{}}SDID Estimate,\\ Relative to Baseline\end{tabular} & 0.1139*** & 0.1141*** \\
                     & (0.0117) & (0.0086)    \\
Country FE           & X        & X   \\
Week FE              & X        & X   \\
Time-Varying Covariates             &          & X   \\ \bottomrule
\multicolumn{3}{l}{\rule{0pt}{1em}* p $<$ 0.1, ** p $<$ 0.05, *** p $<$ 0.01}\\

\end{tabular}%
}
\caption{SDID Results--Dependent Variable: Cryptocurrency outflows}\label{tab:sdid_results}
\end{table}

I show the results of running the SDID procedure in Table \ref{tab:sdid_results}. Since the SDID procedure is run using the level of cryptocurrency outflows, measured in USD, I divide estimates by the pre-treatment mean for the US for better comparability to the main results. Column (1) shows that SDID produces a very similar estimate to that of Poisson QMLE, around an 11.4\% increase in cryptocurrency flows relative to the pre-treatment mean. When I add the pattern of cryptocurrency outflows as covariates, as shown in Column (2), the result is largely unchanged.

\begin{figure}[!h]
    \centering
    \begin{subfigure}{0.49\textwidth}
        \includegraphics[width = \textwidth]{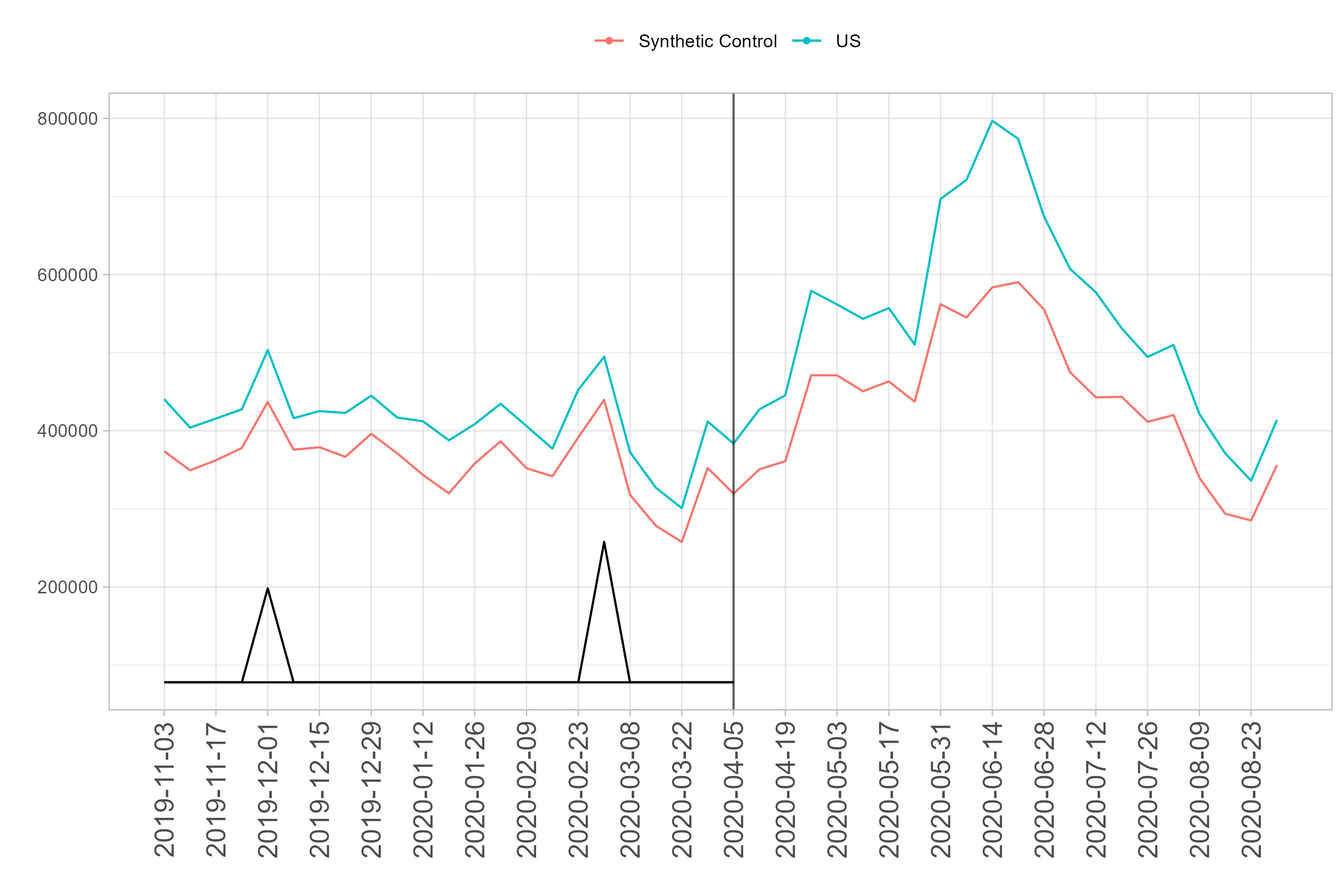}
    \end{subfigure}
    \begin{subfigure}{0.49\textwidth}
        \includegraphics[width = \textwidth]{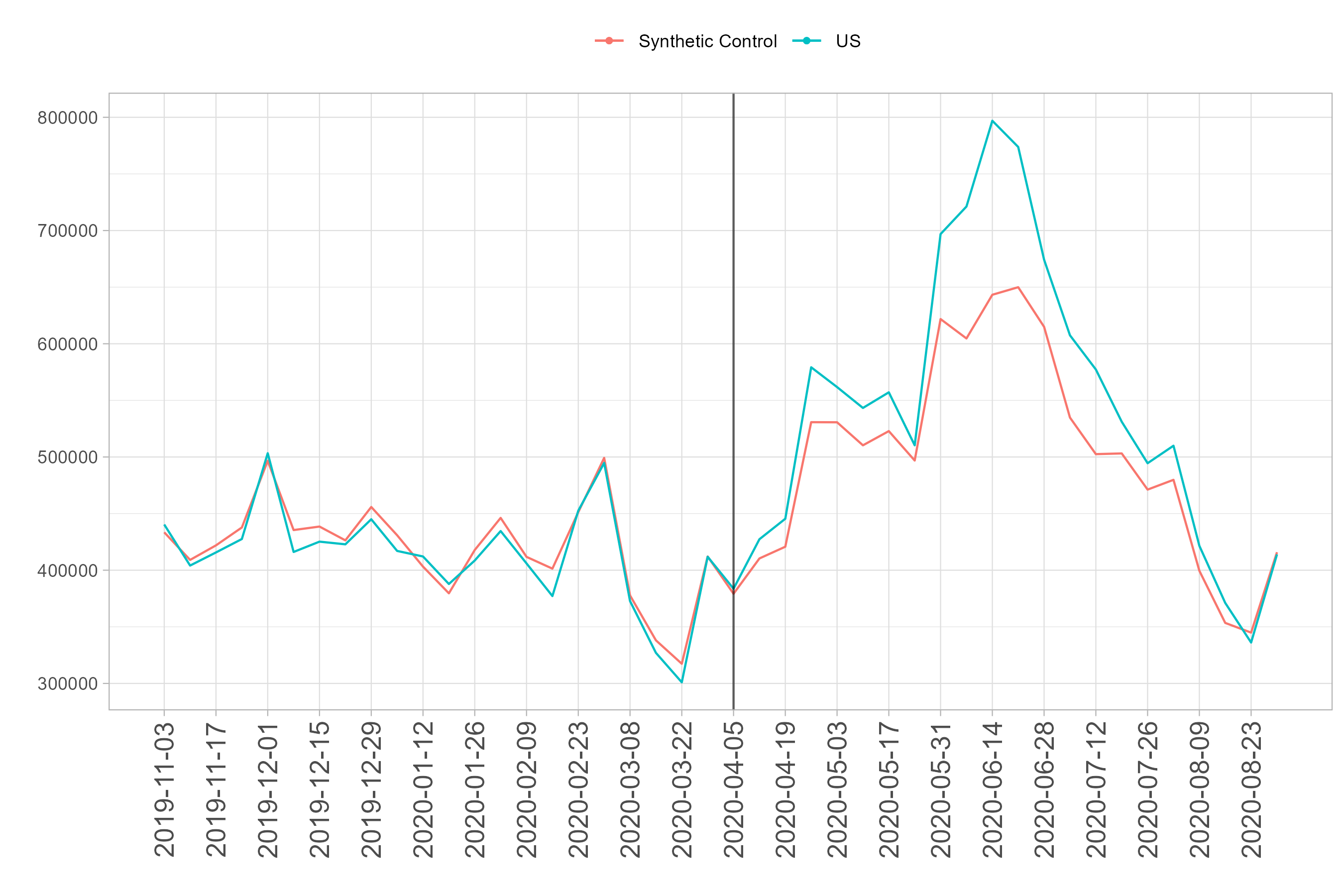}
    \end{subfigure}
    \caption{Synthetic Difference-in-Differences}\label{fig:sdid_main_plots}
\end{figure}

Figures \ref{fig:sdid_main_plots} and \ref{fig:sdid_main_effects_over_time} show the results of estimating the SDID procedure. The left panel of Figure \ref{fig:sdid_main_plots} shows the trajectories of cryptocurrency outflows from the US and the synthetic control, while the right panel shows the same trajectories, but overlays the synthetic control trajectory onto the US, for easier visualization. Figure \ref{fig:sdid_main_effects_over_time} shows the estimated treatment effect over time along with confidence intervals estimated via the placebo procedure. As shown in both figures, cryptocurrency outflows for the US and the synthetic control follow each other closely in the pre-treatment period, then diverge after the disbursement of EIPs. The effects are largest in the first two months after the disbursement of EIPs, then largely dissipate by the end of June 2020, consistent with my main analysis.

\begin{figure}[!h]
    \centering
        \includegraphics[width = 0.8\textwidth]{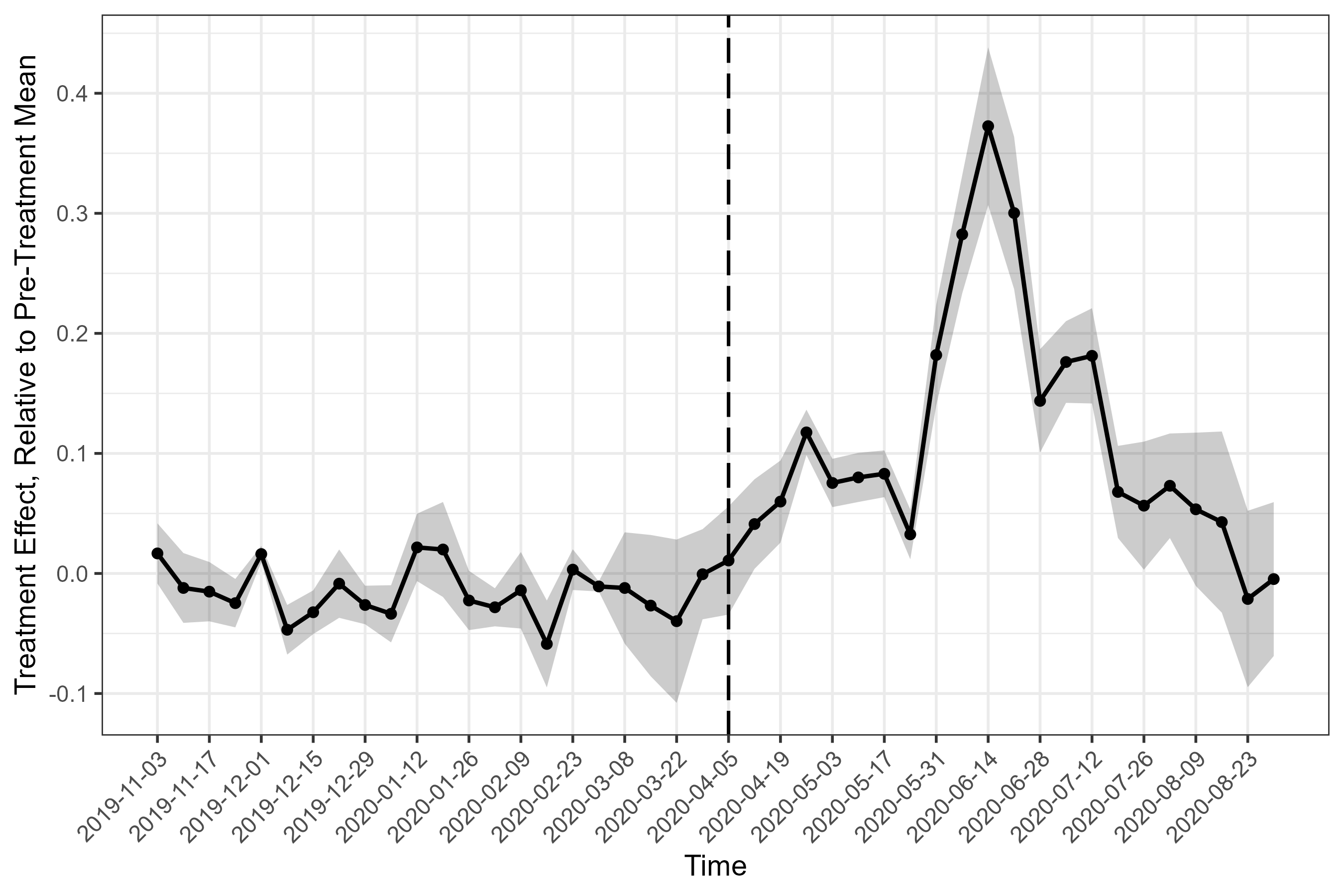}
    \caption{SDID Treatment Effect over Time}\label{fig:sdid_main_effects_over_time}
\end{figure}

\begin{table}[!h]
\begin{tabular}{ll}
Country   & Weight \\ \hline
Nigeria   & 0.679  \\
Ghana     & 0.086  \\
Hong Kong & 0.072  \\
India     & 0.043  \\
United Kingdom & 0.024 \\
\end{tabular}
\caption{SDID Unit Weights}\label{tab:control_weights}
\end{table}

I show the control unit weights in Table \ref{tab:control_weights}, for the estimation including time-varying covariates. The SDID procedure seems to weight control units that are relatively active in terms of cryptocurrency, with Nigeria and Ghana, both known as relatively active countries in terms of cryptocurrency usage, having the two highest weights. Nigeria is by far assigned the highest weight, with a weight around 0.68. As neither of these countries are included in the control group in my main analysis, the fact that the SDID produces similar estimates even with a vastly different control group adds robustness to my findings.

\subsubsection{Placebo in Time}

To add additional robustness to the SDID methodology, I perform two additional placebo tests, this time setting the treatment to a placebo date. In the first test I set the treatment start to January 1st, 2020 and in the other, to exactly two years prior to the actual treatment--far before the beginning of COVID but at the same time of year. In either case, a large placebo estimate when the treatment is set to a placebo date, when no treatment actually occurs, would undermine confidence in the main SDID results.

\begin{figure}[!h]
    \centering
    \begin{subfigure}{0.49\textwidth}
        \includegraphics[width = \textwidth]{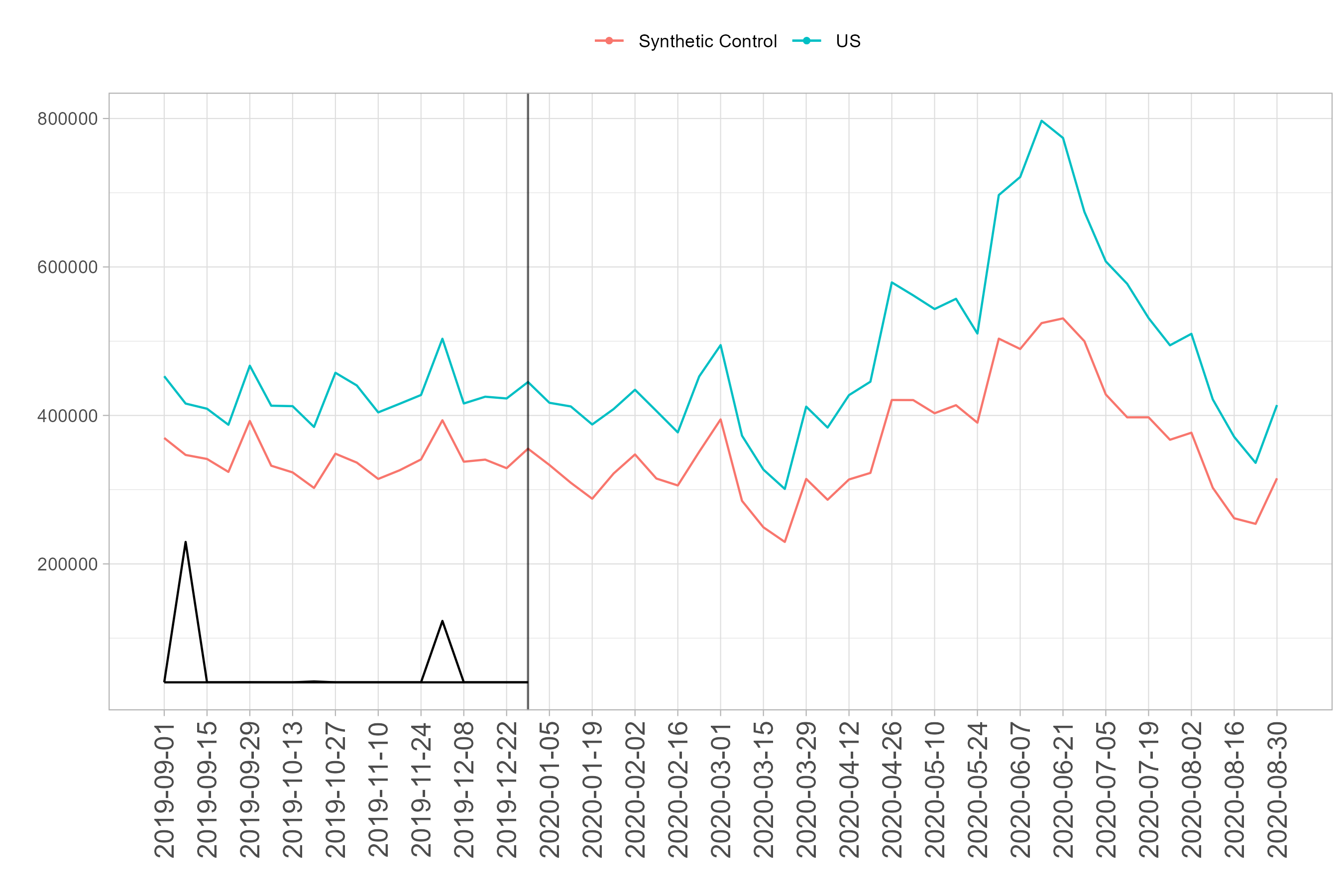}
    \end{subfigure}
    \begin{subfigure}{0.49\textwidth}
        \includegraphics[width = \textwidth]{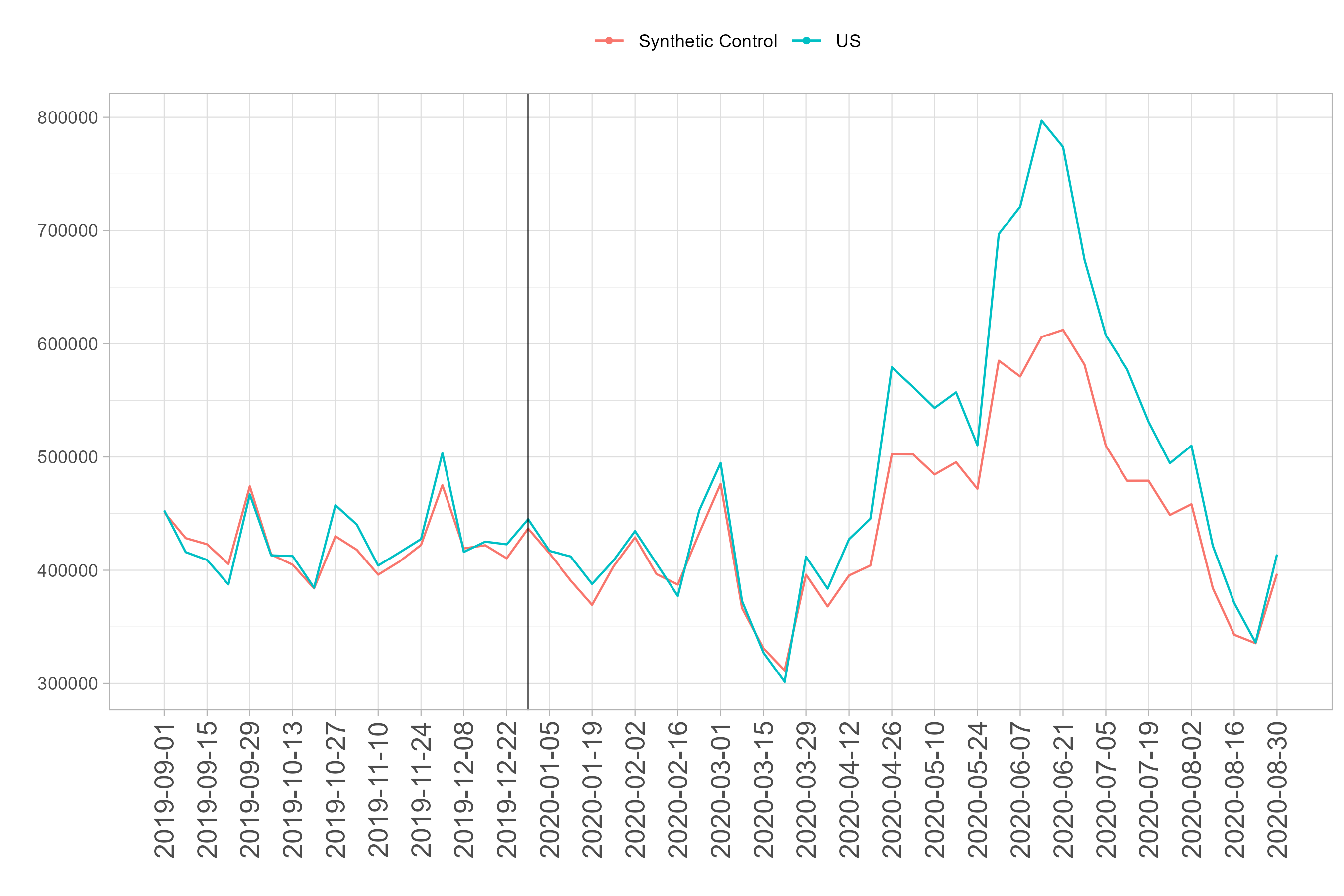}
    \end{subfigure}
    \caption{Jan 2020 Placebo}\label{fig:time_placebo_2020}
\end{figure}

\begin{figure}[!h]
    \centering
    \begin{subfigure}{0.49\textwidth}
        \includegraphics[width = \textwidth]{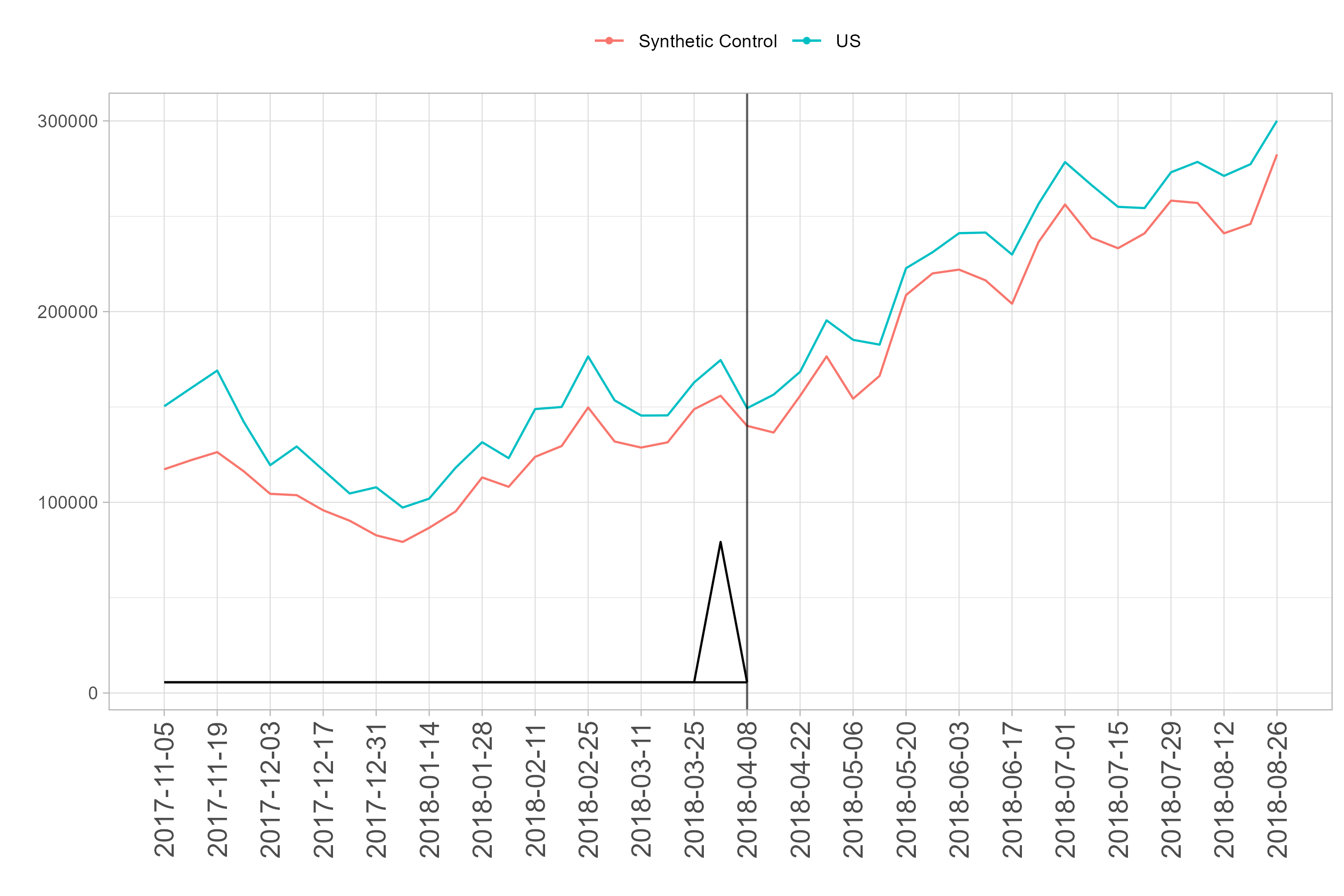}
    \end{subfigure}
    \begin{subfigure}{0.49\textwidth}
        \includegraphics[width = \textwidth]{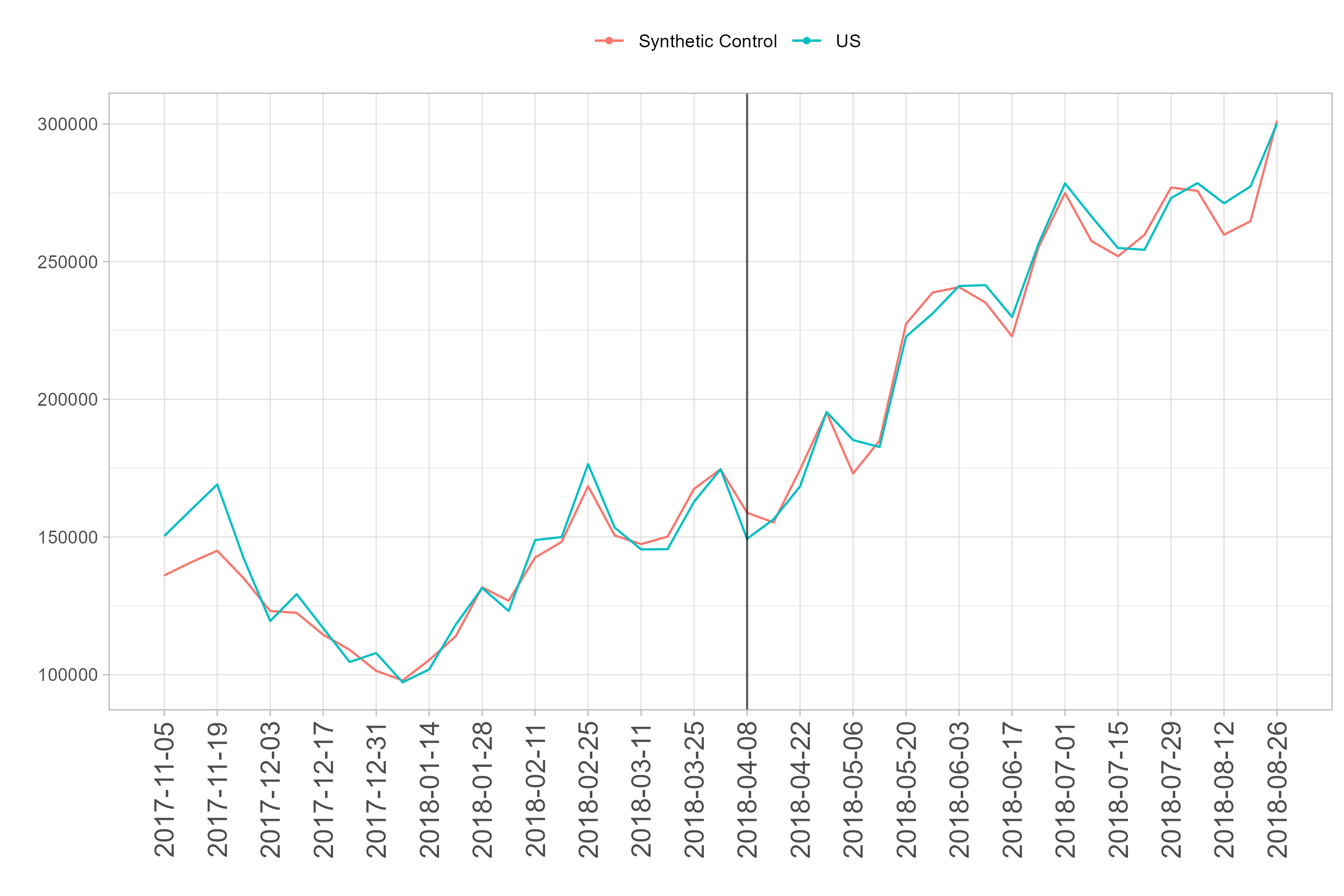}
    \end{subfigure}
    \caption{April 2018 Placebo}\label{fig:time_placebo_2018}
\end{figure}

The results of the time-based placebo tests are shown in Figures \ref{fig:time_placebo_2020} and \ref{fig:time_placebo_2018}. As before, the left panel of each figure shows the true level of outflows of the US and the synthetic control, while the right panel overlays synthetic control trajectory onto that of the US for easier visualization. In the case where treatment is artificially set to the first week of January 2020, as shown in Figure \ref{fig:time_placebo_2020}, the cryptocurrency outflows of the US and synthetic control closely follow each other in the pre-treatment period, but also continue to closely follow one another in the post-treatment period up until April 2020, when stimulus checks are first disbursed, after which they diverge in a way consistent with the main results. In Figure \ref{fig:time_placebo_2018}, when the treatment is artificially set to the first week of April 2018, the trajectory of cryptocurrency outflows between the US and the synthetic control follow each other closely in the entire pre-treatment period \textit{and} the entire post-treatment period, indicating a null effect. As we would not expect to see any effect during this period, this adds robustness to the effect found in 2020.

To summarize, this estimation exercise provides additional robustness to my main results, as the estimated impact of the disbursement of EIPs on cryptocurrency outflows from the US is very close to that found in Section \ref{sect:did_results}, despite the synthetic control being constructed from control units that are generally not included in the control group in my main DID analysis. I additionally find similar dynamics in cryptocurrency outflows over time, with a sharp but temporary increase in cryptocurrency outflows in the two months after disbursement begins. I use SDID as a robustness exercise instead of as my main methodology given the heavy weighting it gives to a select few control units, with this weighting being sensitive to the units included in the donor pool.\footnote{In particular, when I exclude Nigeria from the pool of donor units, the weighting collapses to 100\% Ghana. When I only include high-income countries as the pool of potential donor units, the United Kingdom is given a 100\% weight.} Additionally, the interpretability of the Poisson QMLE estimator as a percentage change is preferable.

\section{Conclusion}\label{sect:conclusion}

In this paper I investigated the effect of fiscal policy on financial outflows. Given the challenges of using traditional macroeconomic data on international financial flows to identify fiscal spillovers, I utilize a granular, detailed dataset on cryptocurrency transactions from the third party cryptocurrency exchange Paxful. Using transaction-level data, I use an algorithm to probabilistically match cryptocurrency transactions to discover ``crypto-vehicle" trades, whereby cryptocurrency is used as a medium for non-speculative financial transactions. Using information on the geolocation of users, I identify when cryptocurrency is sent across borders and construct bilateral cryptocurrency flows at a granular level, allowing for much better identification around fiscal policy shocks than was previously possible.

My main findings can be summarized as follows. Using a difference-in-differences design which exploits the timing of the disbursement of EIPs and compares the US to other high-income countries, I find evidence of a sharp but temporary increase in cryptocurrency outflows associated with the stimulus, with effects largely subsiding around the end of June 2020. I find evidence of even greater increases when focusing on cryptocurrency destined to middle- and high-income destinations, but no effect for cryptocurrency destined to low-income destinations. I then perform a quantification exercise where, using counterfactual US cryptocurrency flows for the entire year of 2020, I calculate the size of the overall impact of the stimulus on cryptocurrency outflows from the US in 2020. This allows me to perform a back-of-the-envelope calculation of the fiscal spillover relative to total expenditure under varying assumptions about the generalizability of my findings. I find the plausible upper bound for the fiscal spillover to be around 2.5\% of the spending allocated towards EIPs in the CARES Act, with more likely estimates between 0.2-0.7\%.

As the size of fiscal spillovers relative to expenditure implied by these estimates are modest, remittances are unlikely to significantly dampen the effectiveness of fiscal policy actions in stimulating output. In an emergency like COVID-19, policymakers can thus prioritize speed in the disbursement of direct stimulus rather than targeting stimulus towards those who are unlikely to remit. It is worth noting, though, that the size of fiscal spillovers could still be large for remittance-receiving countries. The extent to which this is desirable depends on the objectives of policymakers in these countries and current economic conditions. In addition to \cite{decrypting2023}, this study finds evidence that cryptocurrency is used for cross-border money transfers, consistent with suggestions that, relative to traditional remittance channels, cryptocurrency offers advantages in terms of lower fees and faster settlement. For countries reliant on remittances, regulations that improve linkages between cryptocurrency and the traditional financial system or infrastructure supporting internet connectivity could encourage usage of this relatively new financial channel.

While this study provides evidence of fiscal spillovers operating through an informal remittance channel, it requires assumptions about the generalizability of its findings to remittance flows as a whole and the cryptocurrency market in order to calculate overall fiscal spillovers, and while \cite{kpodar_defying_2023} and \cite{bansak2025} document a quick recovery in formal remittances in 2020, both are limited to the monthly cadence of remittance data collected from central banks. Future research could use proprietary data from money operators such as Western Union to explicitly test whether formal remittances see a similar spike in response to the disbursement of EIPs as is found in this study. Additionally, future work can add to the existing literature on fiscal spillovers by studying channels such as the real exchange rate and real interest rate in the context of the COVID-19 pandemic. Finally, as cryptocurrency's role in the global financial market increases over time, more research on its usage as an international financial vehicle is needed to understand how it could shape the international financial landscape going forward.

\clearpage

\bibliographystyle{apalike}
\bibliography{citations}

\clearpage

\appendix
\setcounter{figure}{0}
\setcounter{table}{0}
\renewcommand\thefigure{A.\arabic{figure}}
\renewcommand\thetable{A.\arabic{table}}

\section{Appendix}

\subsection{Transaction Size as Outcome Variable}

In this section I analyze the impact of the disbursement of EIPs on transaction sizes for cross-border cryptocurrency transactions. As previously mentioned, cryptocurrency may offer advantages over traditional financial channels via lower fees, faster settlement, and anonymity. If this is the case, one might expect larger sums to be transferred once the wealth shock of the US COVID stimulus hits. To test this I run a variation of my main analysis in Section \ref{sect:did}, where I use the average weekly transaction size, measured in USD, for cross-border cryptocurrency transactions as the outcome variable. I run the following OLS specification:

\begin{equation}
    TransactionSize_{it} = \beta(Disbursed_{t}\times US_{i}) + \alpha_{i} + \alpha_{t} + \varepsilon_{it}
\end{equation}

where $TransactionSize_{it}$ represents the average transaction size for cross-border cryptocurrency transactions originating from country $i$ in week $t$, measured in USD. As in Equation \ref{eq:qmle_did}, $Disbursed_{t}$ takes on a value of one after EIPs are disbursed on April 9, 2020, and zero otherwise. $US_{i}$ takes on a value of one if country $i$ is the United States, and zero otherwise. $\alpha_{i}$ and $\alpha_{t}$ represent country and time fixed-effects, respectively. $\beta$ thus gives the difference-in-differences estimator, in terms of USD. As in the main analysis, I limit the control group to be countries classified as high-income by the World Bank.

\begin{table}[!h]
\centering
\begin{threeparttable}
\caption{OLS–Dependent Variable: Average Transaction Size (USD)}\label{tab:twfe_transaction_sizes}
\centering
\begin{tabular}[t]{lcccc}
\toprule
  & All Destinations & Low-Income & Middle-Income & High-Income\\
\midrule
$\text{disbursed} \times \text{US}$ & \num{5.429}*** & \num{5.937}*** & \num{4.923}*** & \num{6.472}***\\
 & (\num{0.059}) & (\num{0.120}) & (\num{0.118}) & (\num{0.043})\\
\midrule
$\text{Observations}$ & \num{1357} & \num{621} & \num{1265} & \num{1265}\\
Country FE & X & X & X & X\\
Week FE & X & X & X & X\\
\bottomrule
\multicolumn{5}{l}{\rule{0pt}{1em}* p $<$ 0.1, ** p $<$ 0.05, *** p $<$ 0.01}\\
\end{tabular}
\begin{tablenotes}
\small
\item [a] Standard errors clustered at the country level.
\end{tablenotes}
\end{threeparttable}
\end{table}

Table \ref{tab:twfe_transaction_sizes} shows the results. Similar to the main analysis, I calculate the median transaction size for each country-week pair first with no restriction on the destination of cryptocurrency transactions, then separately for transactions destined to low-, middle-, and high-income countries. In all cases, the disbursement of stimulus checks is associated with an increase in transaction sizes between \$4-\$7, all highly statistically significant. Relative to the average size of all cross-border cryptocurrency transactions in my data, \$46.60, this comes out to between a 8.5-15\% increase in the average cross-border transaction size associated with the stimulus.

\subsection{Inflows as Outcome Variable}\label{sect:inflows}

In this section I use inflows into the US as a ``placebo" test for my main results, given that one would not expect the disbursement of stimulus checks to affect cryptocurrency inflows as it does outflows. I show results of running Equation \ref{eq:qmle_did} with cryptocurrency inflows into the US as the dependent variable, and thus the comparison being inflows into the US compared to inflows into the comparison countries. Following the same reasoning as before, for this analysis I restrict the comparison group to only be other high income countries. Similar to my previous analysis where I restricted the destination of cryptocurrency to countries of varying income groups, in this analysis I run regressions limiting the \textit{origins} of cryptocurrency inflows into the US to be countries from varying income groups. Finally, I use the same period of study as the main analysis.

\begin{table}[!h]
\centering
\begin{threeparttable}
\caption{Poisson QMLE–Dependent Variable: Cryptocurrency Inflows}\label{tab:inflows}
\centering
\begin{tabular}[t]{lcccc}
\toprule
  & All Origins & Low-Income & Middle-Income & High-Income\\
\midrule
$\hat{\theta}_{ATE\%} = e^{\hat{\beta}} - 1$ & \num{0.061} & \num{0.155} & \num{0.167}*** & \num{-0.042}\\
 & (\num{0.050}) & (\num{0.211}) & (\num{0.047}) & (\num{0.066})\\
\midrule
$\text{Observations}$ & \num{1564} & \num{598} & \num{1426} & \num{1426}\\
Country FE & X & X & X & X\\
Week FE & X & X & X & X\\
\bottomrule
\multicolumn{5}{l}{\rule{0pt}{1em}* p $<$ 0.1, ** p $<$ 0.05, *** p $<$ 0.01}\\
\end{tabular}
\begin{tablenotes}
\small
\item [a] Standard errors clustered at the country level.
\end{tablenotes}
\end{threeparttable}
\end{table}

Table \ref{tab:inflows} shows the results. I find a positive increase in inflows into the US following the disbursement of stimulus checks, but only for cryptocurrency flows originating from the middle-income countries. Importantly for the internal validity of my main analysis, where I use high-income countries as a control group, there is no effect for cryptocurrency flows originating from high-income countries. A potential explanation for the increase in inflows from middle-income countries to the US is that while some migrants did receive stimulus checks, many did not. For those that did not and suffered adverse employment effects due to COVID-19, they may have received  financial assistance from their family in their home country. As shown in Figure \ref{fig:migrant_shares}, the top countries in terms of their foreign-born population residing in the US are largely comprised of middle-income countries, thus they may have stronger migrant networks than countries in other income categories.

\setcounter{figure}{0}
\setcounter{table}{0}
\renewcommand\thefigure{B.\arabic{figure}}
\renewcommand\thetable{B.\arabic{table}}

\section{Additional Figures and Tables}

\begin{table}[!h]
\centering
\resizebox{0.6\textwidth}{!}{%
\begin{tabular}{@{}ll@{}}
\toprule
Platform                             & Total 2020 Volume \\ \midrule
\textbf{Paxful}                      & \textbf{\$1.9B}   \\ \midrule
Binance                              & \$1.1T            \\
MEXC                                 & \$626.7B          \\
OKX                                  & \$616.5B          \\
HTX                                  & \$604B            \\
CoinBase                             & \$191.3B          \\
\textbf{Total Cryptocurrency Market} & \textbf{\$44.42T} \\ \midrule
Western Union                        & \$96.1B           \\
Wise                                 & \$76.4B           \\
Remitly                              & \$12.1B           \\
\textbf{Total Official Remittances}  & \textbf{\$717B}   \\ \bottomrule
\end{tabular}%
}
\caption{Comparison of 2020 Volume Across Platforms}\label{tab:volume_comparison}
\floatfoot{Source: cryptocurrency volume retrieved from CoinGecko. Volume for money operators sourced from company financial reports. Overall remittance volume sourced from the World Bank.}
\end{table}

\begin{table}[h]
\centering
\resizebox{\textwidth}{!}{%
\begin{tabular}{lllll}
Trade 1 Timestamp & Trade 1 Country & Trade Size & Trade 2 Timestamp & Trade 2 Country \\ \hline
11/18/2017 6:17   & US              & 0.01283148 & 11/18/2017 7:04   & US              \\
11/18/2017 6:17   & RU              & 0.00104822 & 11/18/2017 6:35   & RU              \\
11/18/2017 6:19   & US              & 0.00115685 & 11/18/2017 6:23   & US              \\
11/18/2017 6:19   & US              & 0.00421585 & 11/18/2017 6:33   & NG              \\
11/18/2017 6:20   & NG              & 0.00219000 & 11/18/2017 8:27   & NG             
\end{tabular}%
}
\caption{Matched Trades Example}
\label{tab:matched_trades}
\end{table}

\begin{figure}[!h]
    \centering
    \caption{Total Cryptocurrency Volume, Broken Down by Category}
    \includegraphics[scale = 0.6]{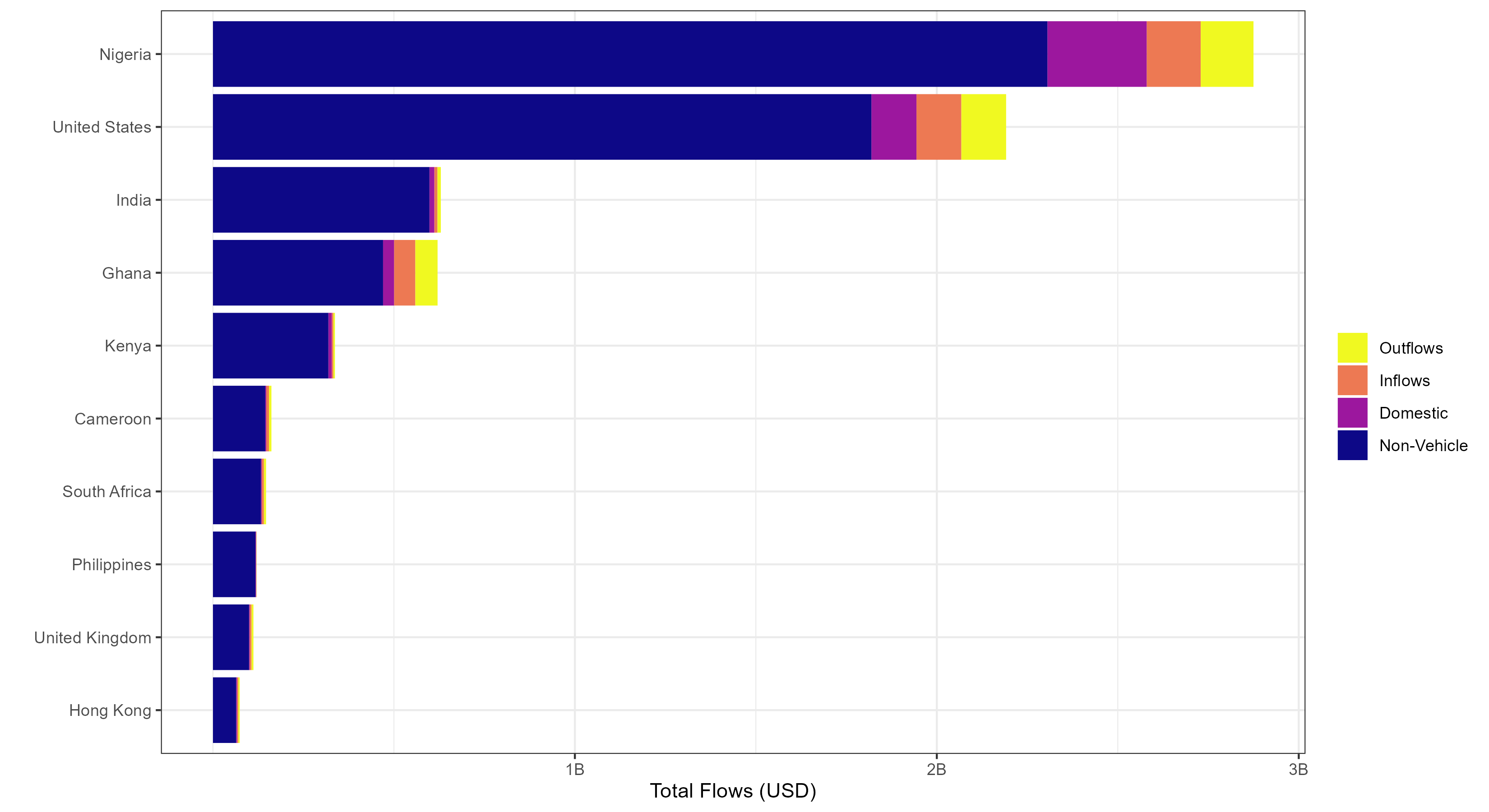}
    \label{fig:flows_by_country}
\end{figure}

\begin{figure}[!h]
    \centering
    \includegraphics[scale = 0.6]{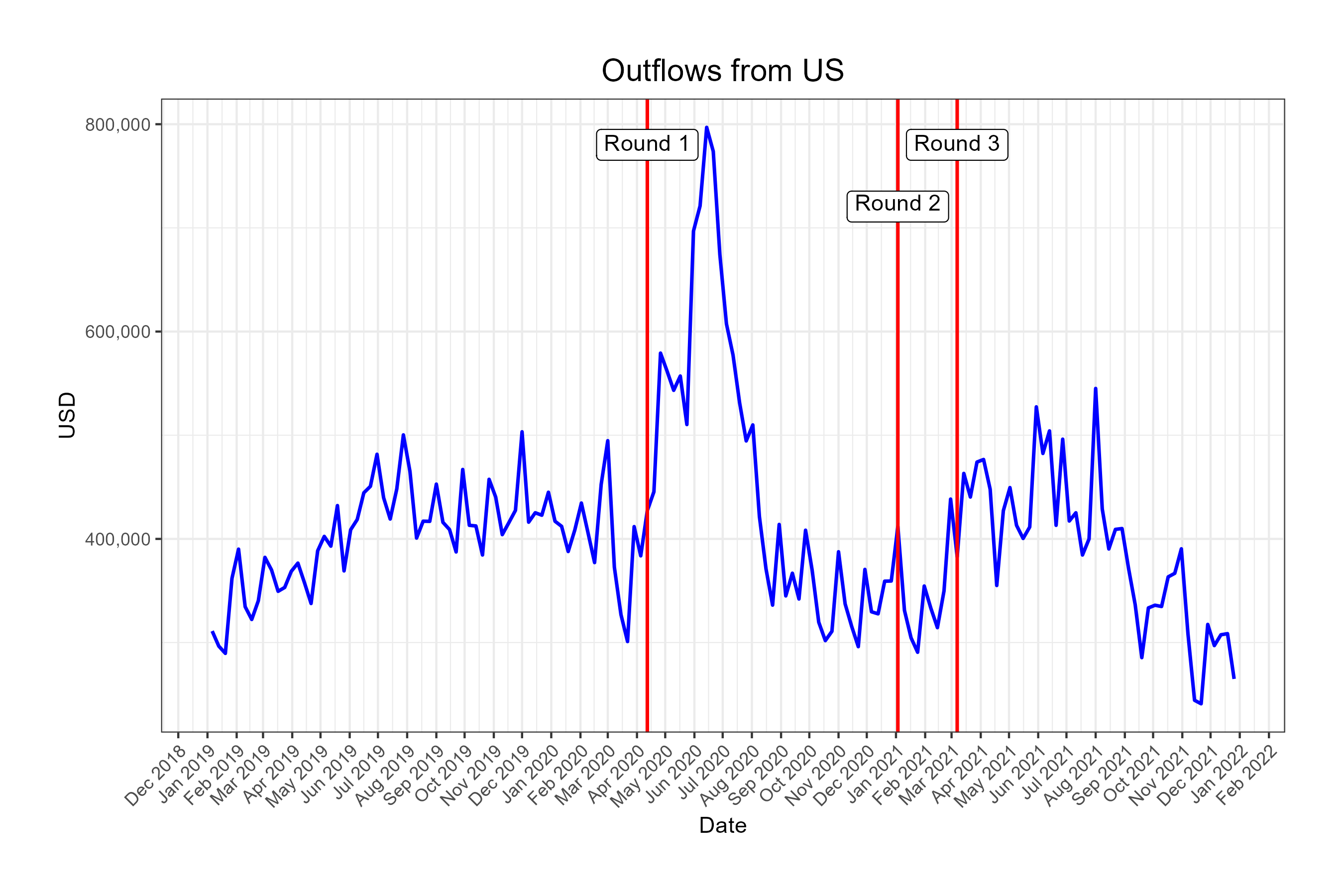}
    \caption{Stimulus Timing, Including 2nd and 3rd Round}
    \floatfoot{Source: Paxful, Author's calculations}
    \label{fig:stimulus_timing_extended}
\end{figure}

\begin{figure}[!h]
    \centering
    \includegraphics[scale = 0.6]{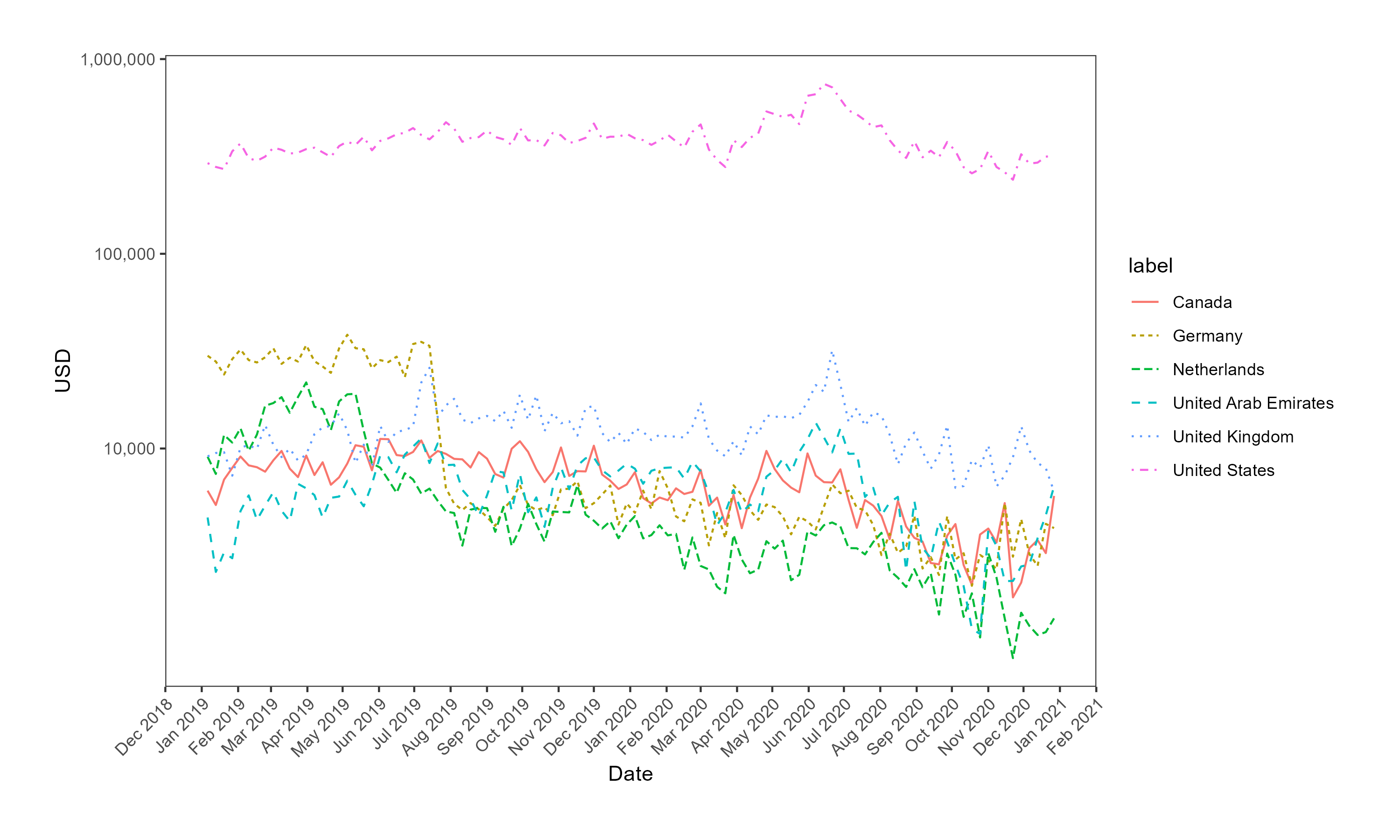}
    \caption{Outflows to Low- and Middle-Income Countries, US vs. Other High Income Countries}
    \label{fig:outflows_us_vs_high_income}
\end{figure}

\begin{table}[]
\resizebox{\textwidth}{!}{
\begin{tabular}{llll}
\caption{World Bank Income Group Classifications}\label{tab:income_classification}
High Income                & Upper-Middle Income              & Lower-Middle Income              & Low-Income                        \\ \hline
Andorra                    & Albania                          & Angola                           & Afghanistan                       \\
United Arab Emirates       & Armenia                          & Bangladesh                       & Burkina Faso                      \\
Antigua and Barbuda        & American Samoa                   & Bolivia (Plurinational State of) & Burundi                           \\
Argentina                  & Azerbaijan                       & Bhutan                           & Benin                             \\
Austria                    & Bosnia and Herzegovina           & Congo                            & Congo, Democratic Republic of the \\
Australia                  & Bulgaria                         & CÃ´te d'Ivoire                   & Eritrea                           \\
Aruba                      & Brazil                           & Cameroon                         & Ethiopia                          \\
Barbados                   & Botswana                         & Cabo Verde                       & Gambia                            \\
Belgium                    & Belarus                          & Djibouti                         & Guinea                            \\
Bahrain                    & Belize                           & Egypt                            & Guinea-Bissau                     \\
Bermuda                    & China                            & Micronesia (Federated States of) & Haiti                             \\
Brunei Darussalam          & Colombia                         & Georgia                          & Comoros                           \\
Bahamas                    & Costa Rica                       & Ghana                            & Liberia                           \\
Canada                     & Cuba                             & Honduras                         & Madagascar                        \\
Switzerland                & Dominica                         & Indonesia                        & Mali                              \\
Chile                      & Dominican Republic               & India                            & Malawi                            \\
CuraÃ§ao                   & Algeria                          & Kenya                            & Mozambique                        \\
Cyprus                     & Ecuador                          & Kyrgyzstan                       & Niger                             \\
Czechia                    & Fiji                             & Cambodia                         & Nepal                             \\
Germany                    & Gabon                            & Kiribati                         & Rwanda                            \\
Denmark                    & Grenada                          & Lao People's Democratic Republic & Sierra Leone                      \\
Estonia                    & Equatorial Guinea                & Sri Lanka                        & Senegal                           \\
Spain                      & Guatemala                        & Lesotho                          & Somalia                           \\
Finland                    & Guyana                           & Morocco                          & South Sudan                       \\
Faroe Islands              & Iraq                             & Moldova, Republic of             & Syrian Arab Republic              \\
France                     & Iran (Islamic Republic of)       & Myanmar                          & Chad                              \\
United Kingdom             & Jamaica                          & Mongolia                         & Togo                              \\
Gibraltar                  & Jordan                           & Mauritania                       & Tajikistan                        \\
Greenland                  & Kazakhstan                       & Nigeria                          & Tanzania, United Republic of      \\
Greece                     & Lebanon                          & Nicaragua                        & Uganda                            \\
Guam                       & Saint Lucia                      & Papua New Guinea                 & Yemen                             \\
Hong Kong                  & Libya                            & Philippines                      & Zimbabwe                          \\
Croatia                    & Montenegro                       & Pakistan                         &                                   \\
Hungary                    & Marshall Islands                 & Palestine, State of              &                                   \\
Ireland                    & North Macedonia                  & Solomon Islands                  &                                   \\
Israel                     & Mauritius                        & Sudan                            &                                   \\
Isle of Man                & Maldives                         & Sao Tome and Principe            &                                   \\
Iceland                    & Mexico                           & El Salvador                      &                                   \\
Italy                      & Malaysia                         & Eswatini                         &                                   \\
Japan                      & Nauru                            & Timor-Leste                      &                                   \\
Saint Kitts and Nevis      & Peru                             & Tunisia                          &                                   \\
South Korea                & Paraguay                         & Ukraine                          &                                   \\
Kuwait                     & Romania                          & Uzbekistan                       &                                   \\
Cayman Islands             & Serbia                           & Viet Nam                         &                                   \\
Liechtenstein              & Russian Federation               & Vanuatu                          &                                   \\
Lithuania                  & Suriname                         & Zambia                           &                                   \\
Luxembourg                 & Thailand                         &                                  &                                   \\
Latvia                     & Turkmenistan                     &                                  &                                   \\
Monaco                     & Tonga                            &                                  &                                   \\
Saint Martin (French part) & Turkey                           &                                  &                                   \\
Macao                      & Tuvalu                           &                                  &                                   \\
Northern Mariana Islands   & Saint Vincent and the Grenadines &                                  &                                   \\
Malta                      & Venezuela                        &                                  &                                   \\
New Caledonia              & Samoa                            &                                  &                                   \\
Netherlands                & South Africa                     &                                  &                                   \\
Norway                     & Namibia                          &                                  &                                   \\
New Zealand                &                                  &                                  &                                   \\
Oman                       &                                  &                                  &                                   \\
Panama                     &                                  &                                  &                                   \\
French Polynesia           &                                  &                                  &                                   \\
Poland                     &                                  &                                  &                                   \\
Puerto Rico                &                                  &                                  &                                   \\
Portugal                   &                                  &                                  &                                   \\
Palau                      &                                  &                                  &                                   \\
Qatar                      &                                  &                                  &                                   \\
Saudi Arabia               &                                  &                                  &                                   \\
Seychelles                 &                                  &                                  &                                   \\
Sweden                     &                                  &                                  &                                   \\
Singapore                  &                                  &                                  &                                   \\
Slovenia                   &                                  &                                  &                                   \\
San Marino                 &                                  &                                  &                                   \\
Sint Maarten (Dutch part)  &                                  &                                  &                                   \\
Turks and Caicos Islands   &                                  &                                  &                                   \\
Trinidad and Tobago        &                                  &                                  &                                   \\
Taiwan, Province of China  &                                  &                                  &                                   \\
United States              &                                  &                                  &                                   \\
Uruguay                    &                                  &                                  &                                   \\
Virgin Islands (British)   &                                  &                                  &                                   \\
Virgin Islands (U.S.)      &                                  &                                  &                                   \\
Virgin Islands (British)   &                                  &                                  &                                   \\
Virgin Islands (U.S.)      &                                  &                                  &                                  
\end{tabular}
}
\end{table}

\begin{figure}[!h]
    \centering
    \begin{subfigure}{.49\textwidth}
    \includegraphics[width = \textwidth]{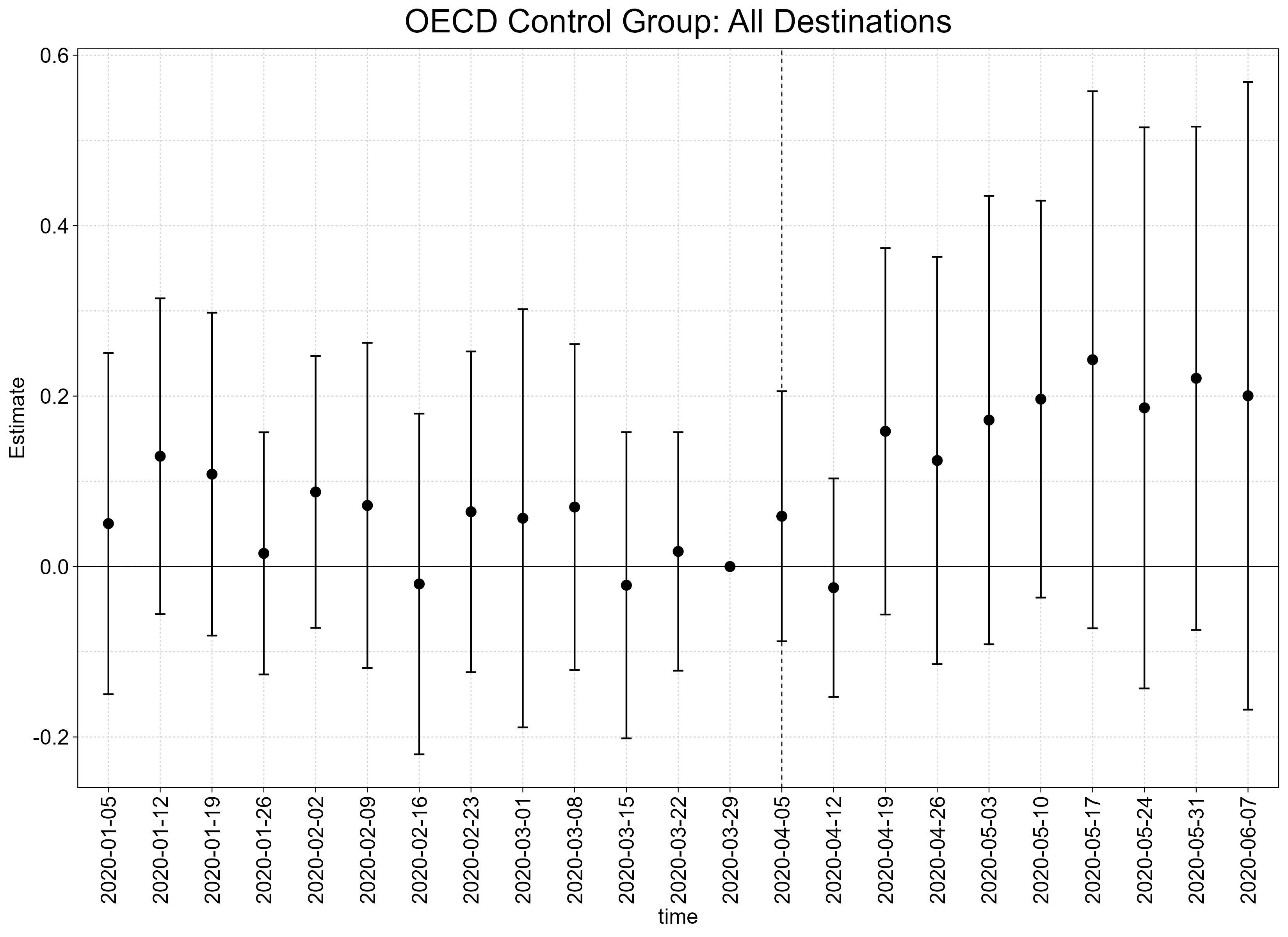}
    \end{subfigure}
    \begin{subfigure}{.49\textwidth}
    \includegraphics[width = \textwidth]{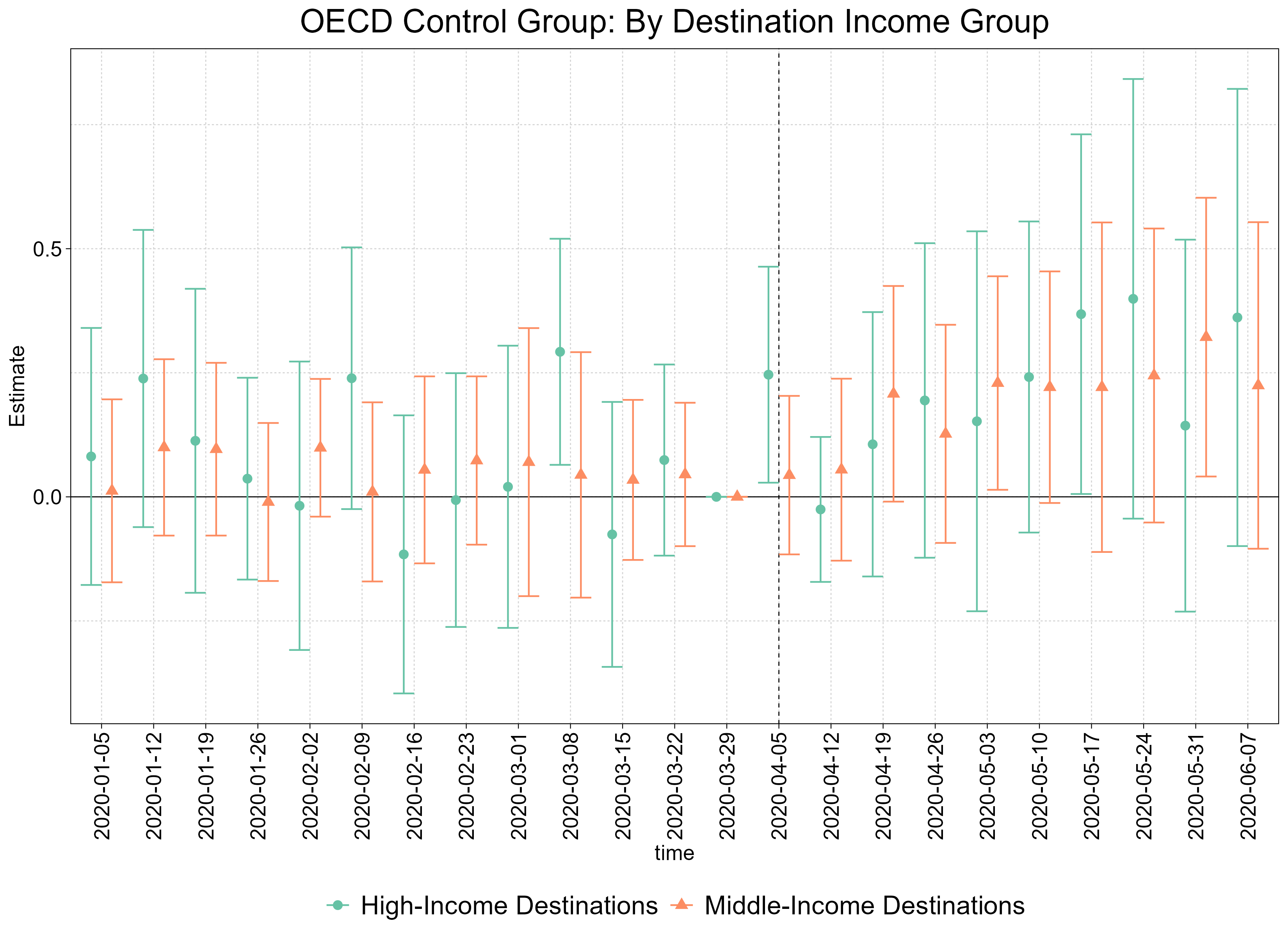}
    \end{subfigure}
    \caption{Event Study Plots: OECD Control Group}
    \label{fig:es_oecd}
\end{figure}

\begin{figure}[!h]
    \centering
    \includegraphics[width=0.55\textwidth]{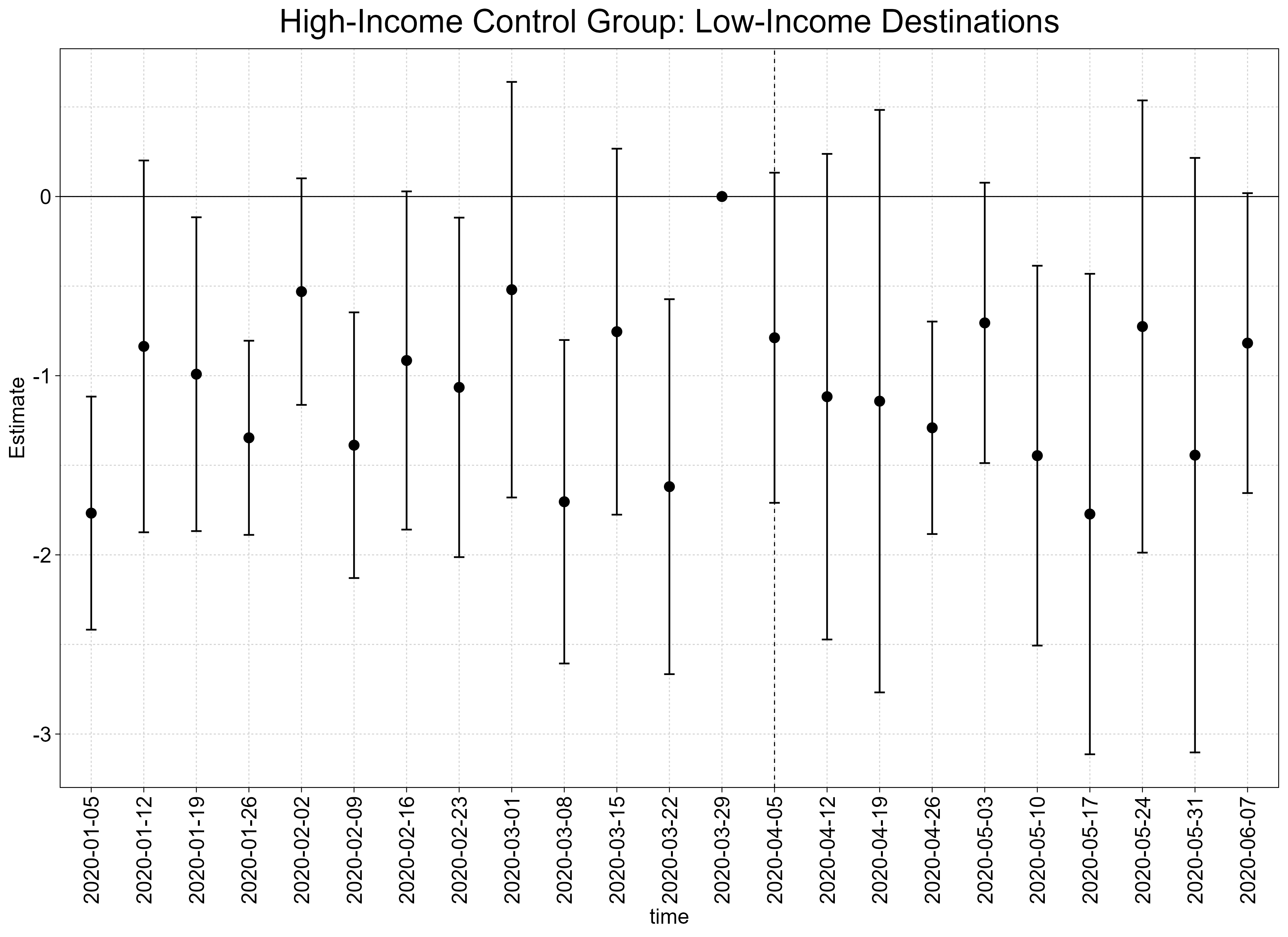}
    \caption{Event Study Plot: Low Income Destinations}
    \label{fig:es_lowincome}
\end{figure}

\begin{table}
\centering
\begin{threeparttable}
\caption{Poisson QMLE–Dependent Variable: Cryptocurrency Outflows}\label{tab:btc_price}
\centering
\begin{tabular}[t]{lcccc}
\toprule
  & All Destinations & Low-Income & Middle-Income & High-Income\\
\midrule
$\text{disbursed} \times \text{US}$ & \num{0.114}*** & \num{-0.072} & \num{0.178}*** & \num{0.159}***\\
 & (\num{0.046}) & (\num{0.238}) & (\num{0.053}) & (\num{0.051})\\
$\text{disbursed}$ & \num{0.181}*** & \num{0.309} & \num{0.106}** & \num{0.300}***\\
 & (\num{0.049}) & (\num{0.335}) & (\num{0.050}) & (\num{0.053})\\
ln(USD/BTC) & \num{1.042}*** & \num{2.015}*** & \num{1.084}*** & \num{0.767}***\\
 & (\num{0.081}) & (\num{0.187}) & (\num{0.043}) & (\num{0.230})\\
\midrule
$\text{Observations}$ & \num{1357} & \num{621} & \num{1265} & \num{1265}\\
Country FE & X & X & X & X\\
\bottomrule
\multicolumn{5}{l}{\rule{0pt}{1em}* p $<$ 0.1, ** p $<$ 0.05, *** p $<$ 0.01}\\
\end{tabular}
\begin{tablenotes}
\small
\item [a] Standard errors clustered at the country level. Bitcoin price is calculated as the average weekly exchange rate between USD and Bitcoin. All estimates transformed via $e^{\beta} - 1.$
\end{tablenotes}
\end{threeparttable}
\end{table}

\end{document}